\documentclass[journal,12pt,onecolumn,draftclsnofoot,]{IEEEtran}

\usepackage[T1]{fontenc}
\usepackage{subcaption}
\usepackage{color}
\usepackage{varioref}
\usepackage{textcomp}
\usepackage{amsthm}
\usepackage{amsmath}
\usepackage{amssymb}
\usepackage{dsfont}
\usepackage{graphicx}
\usepackage{setspace}
\usepackage{esint}
\usepackage{algorithm}
\usepackage{algpseudocode}
\usepackage{pifont}
\usepackage{wrapfig}
\usepackage{multirow}
\usepackage{tabu}
\usepackage{amsmath}

\usepackage{enumitem}
\usepackage{cite}
\usepackage{cleveref}
\makeatletter

\newcommand{\Rmnum}[1]{\expandafter\@slowromancap\romannumeral #1@}
\makeatother

\newtheorem{remark}{Remark}

\theoremstyle{definition}

\providecommand{\propositionname}{Proposition}

\ifCLASSINFOpdf

\else

\fi

\usepackage[T1]{fontenc}
\usepackage{etoolbox}

\makeatletter
\patchcmd{\maketitle}{\@fnsymbol}{\@alph}{}{}  
\makeatother
\title{Federated Learning over Wireless\\ Fading Channels}

\author{\IEEEauthorblockN{Mohammad Mohammadi Amiri and\thanks{M. Mohammadi Amiri is with the Department of Electrical Engineering, Princeton University, Princeton, NJ 08544, USA (e-mail: mamiri@princeton.edu).}\thanks{D. G\"und\"uz is with the Department of Electrical and Electronic Engineering, Imperial College London, London SW7 2AZ, U.K. (e-mail: d.gunduz@imperial.ac.uk).}
Deniz G\"und\"uz}
}
\date{}

\begin{document}

\maketitle

\begin{abstract}
We study federated machine learning at the wireless network edge, where limited power wireless devices, each with its own dataset, build a joint model with the help of a remote parameter server (PS). We consider a bandwidth-limited fading multiple access channel (MAC) from the wireless devices to the PS, and propose various techniques to implement distributed stochastic gradient descent (DSGD) over this shared noisy wireless channel. We first propose a digital DSGD (D-DSGD) scheme, in which one device is selected opportunistically for transmission at each iteration based on the channel conditions; the scheduled device quantizes its gradient estimate to a finite number of bits imposed by the channel condition, and transmits these bits to the PS in a reliable manner. Next, motivated by the additive nature of the wireless MAC, we propose a novel analog communication scheme, referred to as the \textit{compressed analog} DSGD (CA-DSGD), where the devices first sparsify their gradient estimates while accumulating error from previous iterations, and project the resultant sparse vector into a low-dimensional vector for bandwidth reduction. We also design a power allocation scheme to align the received gradient vectors at the PS in an efficient manner. Numerical results show that D-DSGD outperforms other digital approaches in the literature; however, in general the proposed CA-DSGD algorithm converges faster than the D-DSGD scheme and other schemes in the literature, and reaches a higher level of accuracy. We have observed that the gap between the analog and digital schemes increases when the datasets of devices are not independent and identically distributed (i.i.d.). Furthermore, the performance of the CA-DSGD scheme is shown to be robust against imperfect channel state information (CSI) at the devices.
Overall these results show clear advantages for the proposed analog over-the-air DSGD scheme, which suggests that learning and communication algorithms should be designed jointly to achieve the best end-to-end performance in machine learning applications at the wireless edge.\makeatletter{\renewcommand*{\@makefnmark}{}\footnotetext{This work was supported in part by the European Research Council (ERC) Starting Grant BEACON (grant agreement no. 725731).}\makeatother}
\end{abstract}


\section{Introduction}\label{SecIntro}


As the dataset sizes and model complexities grow, distributed machine learning (ML) is becoming the only viable alternative to centralized ML. In particular, with the increasing amount of information collected through wireless edge devices, such centralized solutions are becoming increasingly costly, due to the limited power and bandwidth available, and less desirable due to privacy concerns. Federated learning (FL) has been proposed as an alternative privacy-preserving distributed ML scheme, where each device participates in training using only locally available data with the help of a parameter server (PS) \cite{DCKonecnyFederated}. In FL devices exchange model parameters and their local updates with the PS, but the data never leaves the devices. As mentioned, in addition to privacy benefits, this is an attractive approach for wireless edge devices when dataset sizes are very large.

ML problems often involve the minimization of the empirical loss function 
\begin{align}\label{EmpLossFunc}
F \left( \boldsymbol{\theta} \right) = \frac{1}{\left| \mathcal{B} \right|} \sum\limits_{\boldsymbol{u} \in \mathcal{B}} f \left(\boldsymbol{\theta}, \boldsymbol{u} \right),
\end{align}
where $\boldsymbol{\theta} \in \mathbb{R}^d$ denotes the model parameters to be optimized, $\cal B$ is the training dataset of size $\left| \mathcal{B} \right|$ consisting of data samples and their labels, and $f(\cdot)$ is the loss function defined by the learning task. 
The minimization of $F \left( \boldsymbol{\theta} \right)$ is typically carried out through iterative stochastic gradient descent (SGD) algorithm, in which the model parameter vector at iteration $t$, $\boldsymbol{\theta}_t$, is updated with a stochastic gradient
\begin{align}\label{SGDModelUpdateBasic}
\boldsymbol{\theta}_{t+1} =\boldsymbol{\theta}_{t} - \eta_t \boldsymbol{g} \left(\boldsymbol{\theta}_{t} \right),
\end{align}
which satisfies $\mathbb{E} \left[ \boldsymbol{g} \left(\boldsymbol{\theta}_{t} \right) \right] = \nabla F \left( \boldsymbol{\theta}_t \right)$, where $\eta_t$ is the learning rate. SGD can easily be implemented across multiple devices, each of which has access to only a small fraction of the dataset. In distributed SGD (DSGD), at each iteration, device $m$ computes a gradient vector based on the global parameter vector with respect to its local dataset, denoted by $\mathcal{B}_{m}$, and sends the result to the PS, which updates the global parameter vector according to
\begin{align}\label{ParallelSGDModelUpdate}
\boldsymbol{\theta}_{t+1} =\boldsymbol{\theta}_{t} - \eta_t \frac{1}{M} \sum\nolimits_{m=1}^{M} \boldsymbol{g}_m \left(\boldsymbol{\theta}_{t} \right),   
\end{align}
where $M$ denotes the number of wireless devices, and $\boldsymbol{g}_m \left(\boldsymbol{\theta}_{t} \right) \triangleq \frac{1}{\left| \mathcal{B}_{m} \right|} \sum\nolimits_{\boldsymbol{u} \in \mathcal{B}_{m}} \nabla f \left(\boldsymbol{\theta}_t, \boldsymbol{u} \right)$, $m \in [M]$. In FL, each device participating in the training can also carry out multiple model updates as in \eqref{ParallelSGDModelUpdate} locally, and share the overall difference with respect to the previous global model parameters with the PS \cite{DCKonecnyFederated}. 

What distinguishes FL from conventional ML is the large number of devices that participate in the training, and the low-capacity and unreliable links that connect these devices to the PS. Therefore, there have been significant research efforts to reduce the communication requirements in FL \cite{DCKonecnyFederated,GoogleMcMahanFed,KonecnyRandDistMean,McMahan2017CommunicationEfficientLO,SmithFedMultiTask,DCLimitedPrecisionGupta,DCOneBitQuan,DCAlistarhQSGD,DCWenTernGrad,DorefaZhou,DCWangATOMO,SignSGDBernstein,HardwareLi,ScalableDNNStorm,DCAjiSparse,DCLinHanDeepGradComp,MePropSun,DCSattlerSparseBinary,SparCMLRenggli,SparseConvergenceAlistarh,VarBasedTsuzuku,LocalSGDStich,UseLocalSGDLin,LAGGradientGiannakis}. However, these and follow-up studies ignore the physical layer aspects of wireless connections and consider interference-and-error-free links from the participating devices to the PS, even though FL has been mainly motivated for mobile devices.



In this paper, we consider DSGD \textit{over-the-air}; that is, we consider a wireless shared medium from the devices to the PS over which they send their gradient estimates. To emphasize the limitations of the wireless medium, we note that the dimension of some of the recent ML models, which also determines the size of the gradient estimates or model updates that must be transmitted to the PS at each iteration, can be extremely large, e.g., the 50-layer ResNet network has $\sim26$ million weight parameters, while the VGGNet architecture has approximately $138$ million parameters. On the other hand, available channel bandwidth is typically small due to the bandwidth and latency limitations; for example 1 LTE frame of 5MHz bandwidth and duration 10ms can carry only 6000 complex symbols.
In principle, we can treat each iteration of the DSGD algorithm as a distributed over-the-air lossy computation problem. FL over a static Gaussian MAC is studied in \cite{MohammadDenizDSGDCS}, where both a digital scheme, which separates computation and communication, and an analog over-the-air computation scheme are introduced. While the digital scheme exploits gradient quantization followed by independent channel coding at the participating wireless devices, the analog scheme exploits the additive nature of the wireless channel and gradient sparsification, and employs random linear projection for dimensionality reduction. In \cite{KaibinParallelWork} the authors consider a fading MAC, and also apply analog transmission, where each entry of a gradient vector at each of the devices is scheduled for transmission depending on the corresponding channel condition. A multi-antenna PS is considered in \cite{YangFedLearOverAirComp}, where receive beamforming is used to maximize the number of devices scheduled for transmission at each iteration.
%


Here, we extend our previous works \cite{MohammadDenizDSGDCS,MohamamdDenizFLOverAirSPAWC19}, and study DSGD over a wireless fading MAC. While we consider gradient descent, where each device sends its local gradient estimate at each iteration, the results can easily be extended by letting the devices send their model updates after several local SGD iterations. We first consider the \textit{separate} computation and communication approach, and propose a digital DSGD (D-DSGD) scheme, in which only a single device is opportunistically scheduled for transmission at each iteration of DSGD based on the channel conditions from the devices to the PS. The scheduled device quantizes its gradient estimate to a finite number of bits using the gradient compression scheme in \cite{DCSattlerSparseBinary} while accumulating the error from previous iterations (this will be clarified later), and employs a channel code to transmit the bits over the available bandwidth-limited channel to the PS. For the MNIST classification task, it is shown that the proposed digital approach D-DSGD outperforms digital schemes that employ QSGD \cite{DCAlistarhQSGD} or SignSGD \cite{SignSGDBernstein} for gradient compression. We also observe that the proposed opportunistic scheduling scheme outperforms the scheme when all the devices participate in the transmission, with each device allocated orthogonal channel resources to communicate with the PS.

We then study analog transmission from the devices to the PS motivated by the signal-superposition property of the wireless MAC. At first, we extend the scheme in \cite{KaibinParallelWork} by introducing error accumulation, which is shown to improve the performance. We then propose a novel scheme, inspired by the random projection used in \cite{MohammadDenizDSGDCS} for dimensionality reduction, which we will refer to as the \textit{compressed analog DSGD} (CA-DSGD). With CA-DSGD, we exploit the similarity in the sparsity patterns of the gradient estimates at different devices to speed up the computations, where each device projects its gradient estimate to a low-dimensional vector and transmits only the important gradient entries while accumulating the error. CA-DSGD scheme provides the flexibility of adjusting the dimension of the gradient estimate sent by each device, which is particularly important for bandwidth-limited wireless channels, where the bandwidth available for transmission may not be sufficient to send the entire gradient vector at a single time slot. A power allocation scheme is also designed, which aligns the vectors sent by different devices at the PS while satisfying the average power constraint. Numerical results for the MNIST classification task show that the proposed CA-DSGD scheme improves upon the other analog and digital schemes under consideration with the same average power constraint and bandwidth resources, with the improvement more significant when the datasets across devices are non-independent and identically distributed (i.i.d.). Its performance is also shown to be robust against imperfect channel state information (CSI) at the devices, whereas digital schemes are sensitive to accurate CSI at the devices, particularly if close to capacity operation is desired. 
In addition to these benefits of the proposed CA-DSGD scheme we make the following observations: 
\begin{enumerate}[label=\arabic*)]
\item The improvement of analog over-the-air computation compared to the D-DSGD scheme is particularly striking in the low power regime. This is mainly due to the ``beamforming'' effect of simultaneously transmitting highly correlated gradient estimates. 
\item While both the convergence speed and the accuracy of the D-DSGD scheme increase significantly with the available average power, the performance of the analog schemes improve marginally. This highlights the energy efficiency of over-the-air computation, and makes it particularly attractive for FL across low-power IoT sensors. 
\item Increasing the number of devices improves the accuracy for all the schemes even if the total dataset size and total power consumption remain the same. This ``diversity gain'' is much more limited for the analog scheme, and diminishes further as the training duration increases. 
\item We observe that the performance of the CA-DSGD scheme improves if we reduce the bandwidth used at each iteration, and increase the number of DSGD iterations instead. 
\end{enumerate}



\textit{Notations}: $\mathbb{R}$ and $\mathbb{C}$ represent the sets of real and complex values, respectively. For vectors $\boldsymbol{x}$ and $\boldsymbol{y}$ with the same dimension, $\boldsymbol{x} \circ \boldsymbol{y}$ returns their Hadamard/entry-wise product. For a vector $\boldsymbol{z} \in \mathbb{C}^i$, ${\rm{Re } \{ \boldsymbol{z} \}} \in \mathbb{R}^i$ and ${\rm{Im } \{ \boldsymbol{z} \}} \in \mathbb{R}^i$ return the entry-wise real and imaginary components of $\boldsymbol{z}$, respectively. Also, $[\boldsymbol{v}, \boldsymbol{w}]$ represents the concatenation of two row vectors $\boldsymbol{v}$ and $\boldsymbol{w}$. We denote a zero-mean normal distribution with variance $\sigma^2$ by $\mathcal{N} \left( 0,\sigma^2 \right)$, and $\mathcal{C N} \left( 0,\sigma^2 \right)$ represents a complex normal distribution with independent real and imaginary terms each distributed according to $\mathcal{N} \left( 0,\sigma^2 / 2 \right)$. For positive integer $i$, we let $[i] \triangleq \{ 1, \dots, i \}$. We denote the cardinality of set $\cal A$ by $\left| \mathcal{A} \right|$, and $l_2$ norm of vector $\boldsymbol{x}$ by $\left\| \boldsymbol{x} \right\|_2$. The imaginary unit is represented by $j$.



\begin{figure}[t!]
\centering
\includegraphics[scale=0.65]{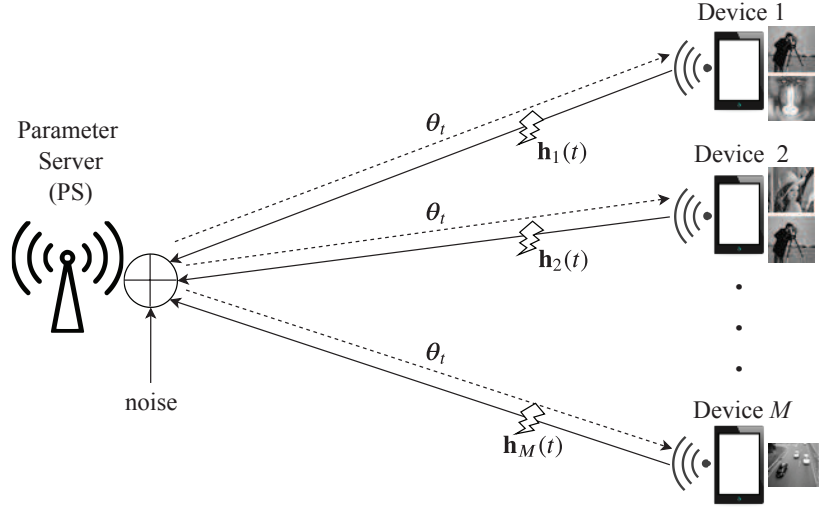}
\caption{Illustration of wireless FL architecture. The PS sends the updated parameter vector to all the wireless devices over an error-free ideal multicast channel, while the gradient estimates, computed by each device using only the available local dataset, are transmitted to the PS over the fading uplink channel.}
\label{System_Model}
\end{figure}

\section{System Model}\label{SecProbFormul}

We consider FL across $M$ wireless devices, each with its own local dataset, which employ DSGD with the help of a remote PS. We model the channel from the devices to the PS as a wireless fading MAC, and OFDM is employed for transmission. The system model is illustrated in Fig. \ref{System_Model}. The parameter vector at iteration $t$ is denoted by $\boldsymbol{\theta}_t$, and we assume that it is delivered from the PS to the devices over an error-free shared link. We denote the set of data samples available at device $m$ by $\mathcal{B}_{m}$, with $\left| \mathcal{B}_m \right| = B$, $\forall m \in [M]$, and the stochastic gradient computed by device $m$ with respect to local data samples by $\boldsymbol{g}_m \left(\boldsymbol{\theta}_{t} \right) \in \mathbb{R}^d$, $m \in [M]$. At the $t$-th iteration of the DSGD algorithm in \eqref{ParallelSGDModelUpdate}, the local gradient estimates of the devices are sent to the PS over a wireless fading MAC using $s$ subchannels for a total of $N$ time slots, where $s \le d$ (in practice, we typically have $s \ll d$). We denote the length-$s$ channel input vector transmitted by device $m$ at the $n$-th time slot of the $t$-th iteration of the DSGD by $\boldsymbol{x}_{m}^n (t) =[x^n_{m,1} (t) \cdots x^n_{m,s} (t)]^T \in \mathbb{C}^s$. The channel output $\boldsymbol{y}^n(t) \in \mathbb{C}^s$ received by the PS at the $n$-th time slot of the $t$-th iteration, $n \in [N]$, is given by
\begin{align}\label{ReceivedVectorPSGen}
\boldsymbol{y}^n (t) = \sum\nolimits_{m =1}^{M} \boldsymbol{h}^n_{m} (t) \circ \boldsymbol{\mu}_m^n(t) \circ \boldsymbol{x}^n_{m} (t) + \boldsymbol{z}^n (t),
\end{align}
where $\boldsymbol{\mu}^n_m (t) \in \{0,1\}^{s}$ is the entry-wise scheduling vector with the $i$-th entry $\mu^n_{m,i}(t) = 1$, if $m \in \mathcal{M}_i^n(t)$, and $\mu^n_{m,i}(t) = 0$, otherwise\footnote{$\mathcal{M}_i^n(t)$ is a subset of the devices that will be specified for each of the schemes.}, $\boldsymbol{h}^n_m (t) \in \mathbb{C}^s$ is the channel gains vector from device $m$ to the PS with the $i$-th entry ${h}^n_{m, i} (t)$ i.i.d. according to $\mathcal{C N} (0, \sigma^2)$, e.g., Rayleigh fading, and $\boldsymbol{z}^n(t) \in \mathbb{C}^s$ is complex Gaussian noise vector with the $i$-th entry $z^n_{i} (t)$ i.i.d. according to $\mathcal{C N} \left( 0, 1 \right)$. The channel input vector of device $m$ at the $n$-th time slot of iteration $t$, $n \in [N]$, is a function of the channel gain vector $\boldsymbol{h}^n_m (t)$, current parameter vector $\boldsymbol{\theta}_{t}$, the local dataset $\mathcal{B}_{m}$, and the current gradient estimate at device $m$, $\boldsymbol{g}_m \left(\boldsymbol{\theta}_{t} \right)$, $ m \in [M]$. We assume that, at each time slot, the CSI is known by the devices and the PS. For a total of $T$ iterations of the DSGD algorithm, the following total average transmit power constraint is imposed at device $m$:
\begin{align}\label{AvePowerConsGen}
\frac{1}{NT} \sum\nolimits_{t=1}^{T} \sum\nolimits_{n=1}^{N} \mathbb{E} \left[ ||\boldsymbol{x}^n_{m} (t)||^2_2 \right] \le \bar{P}, \quad \forall m \in [M],
\end{align}
where the expectation is taken over the randomness of the channel gains.

The goal is to recover $\frac{1}{M} \sum\nolimits_{m=1}^{M} \boldsymbol{g}_m \left(\boldsymbol{\theta}_{t} \right)$ at the PS, which then updates the model parameter as in \eqref{ParallelSGDModelUpdate} after $N$ time slots. However, due to the pre-processing performed at each device and the distortion caused by the wireless channel, the PS uses a noisy estimate to update the model parameter. Having defined $\boldsymbol{y}(t) \triangleq [{\boldsymbol{y}^1 (t)}^T \cdots {\boldsymbol{y}^N (t)}^T]^T$, we have $\boldsymbol{\theta}_{t+1} = \phi (\boldsymbol{\theta}_{t}, \boldsymbol{y} (t))$ for some update function $\phi : \mathbb{R}^d \times \mathbb{C}^{Ns} \to \mathbb{R}^d$. The updated model parameter is then multicast to the devices by the PS through an error-free shared link, so the devices receive a consistent parameter vector for their computations in the next iteration.

We remark that the goal is to recover the average of the local gradient estimates of the devices at the PS, which is a distributed lossy computation problem over a noisy MAC. We will consider both a digital approach based on separating computations and communication, and an analog transmission approach, where the gradients are transmitted simultaneously over the wireless MAC in an analog fashion, without being converted into bits first.
Analog transmission has been well studied for image/ video multicasting over wireless channels in recent years \cite{JakubczakKatabiSoftCast,ImagVideoPowerDistXiong,SparseCastTungDeniz}, and here we employ the random projection technique proposed in \cite{SparseCastTungDeniz} for image transmission over a bandwidth limited wireless channel.

\section{Digital DSGD}\label{SecDigital}
We first consider DSGD with digital transmission of the gradient estimates by the devices over the wireless fading MAC, referred to as the digital DSGD (D-DSGS) scheme. For D-DSGD, we consider $N = 1$, i.e., the parameter vector is updated after each time slot, and drop the dependency on time slot parameter $n$.

The goal here is to schedule devices and employ power allocation across time slots such that devices can transmit to the PS their local gradient estimates as accurately as possible. A possible approach is to schedule all the devices at all the iterations; however, due to the interference among the devices this will result in each device sending a very coarse description of its local gradient estimate. Instead here we will schedule the devices opportunistically according to their channel states. 

In particular, with the knowledge of channel state information (CSI), at each iteration $t$, we select the device with the largest value of $\sum\nolimits_{i=1}^{s} \left| h_{m,i} (t) \right|^2$, $m \in [M]$. Accordingly, the index of the transmitting device at iteration $t$ is given by:
\begin{align}\label{WorkSelectionDig}
m^* (t) = \arg \mathop {\max }\limits_{m \in \left[ M \right]} \left\{ \sum\nolimits_{i=1}^{s} \left| h_{m,i} (t) \right|^2 \right\}. 
\end{align}
We note that, due to the symmetry in the model, the probability of selecting a device at any time is the same, $1/M$. The power allocated to device $m$ at the $t$-th iteration is given by $\bar{P}_m (t)$, where $\bar{P}_m (t)=0$, if $m \ne m^* (t)$, and it should satisfy
\begin{align}\label{AvePowerConsDDSGD}
\frac{1}{MT} \sum\nolimits_{t=1}^{T} \bar{P}_m (t) \le \bar{P}, \quad \mbox{for $m \in [M]$}. 
\end{align}
For the rate of transmission, we will use a capacity upper bound. The $i$-th entry of the channel output at the $t$-th iteration, which is the result of transmission from device $m^* (t)$, is given by
\begin{align}\label{ReceivedVectorPSDDSGD}
y_i (t) = {h}_{m^* (t),i} \left(t\right) {x}_{m^* (t), i} (t) + {z}_i (t), \quad \mbox{$i \in [s]$}, 
\end{align}
which is equivalent to a wireless fast fading channel with a limited number of $s$ channel uses, with CSI known at both the transmitter and the receiver. In the following, we provide an upper bound on the capacity of this channel by treating it as $s$ parallel Gaussian channels. This is equivalent to coding across infinitely many realizations of this $s$-dimensional channel. 
For a transmit power $P_{m^* (t)} (t)$, the capacity of this parallel Gaussian channel is the result of the following optimization problem \cite[Section 5.4.6]{TseWirelessCommun}:
\begin{align}\label{CapacityDDSGD}
&\mathop {\max }\limits_{P_{1}, \dots, P_{s}} \sum\nolimits_{i=1}^{s} {\log _2}\left( 1 + P_{i} \left| {h}_{m^* (t),i} \left(t\right) \right|^2 \right), \nonumber\\
& \mbox{subject to $\sum\nolimits_{i=1}^{s} P_{i} = \bar{P}_{m^* (t)} (t)$}.
\end{align}
The optimization problem in \eqref{CapacityDDSGD} is solved through waterfilling, and the optimal power allocation is given by
\begin{align}\label{CapacityDDSGDWaterFillingPA}
P^*_{i} = \max \left\{ \frac{1}{\zeta} - \frac{1}{\left| {h}_{m^* (t),i} \left(t\right) \right|^2} , 0 \right\},
\end{align}
where $\zeta$ is determined such that $\sum\nolimits_{i=1}^{s} P^*_{i} = \bar{P}_{m^* (t)} (t)$. Having calculated $P^*_{1}, \dots, P^*_{s}$, the capacity of the wireless channel in \eqref{ReceivedVectorPSDDSGD} is given by
\begin{align}\label{CapacityDDSGDRt}
R (t) = \sum\nolimits_{i=1}^{s} {\log _2}\left( 1 + P^*_{i} \left| {h}_{m^* (t),i} \left(t\right) \right|^2 \right),
\end{align}
which provides an upper bound on the capacity of the communication channel between device $m^* (t)$ and the PS. We would like to emphasize that this capacity upper bound can be quite loose especially for small $s$ values.

We adopt the D-DSGD scheme proposed in \cite[Section III]{MohammadDenizDSGDCS}, in which the gradient estimate ${\boldsymbol{g}}_{m} \left( \boldsymbol{\theta}_t \right)$, computed at device $m$, is added to the error accumulated from previous iterations, denoted by $\boldsymbol{\Delta}_{m} (t-1)$, where we set $\boldsymbol{\Delta}_{m} (0) = \boldsymbol{0}$, $m \in [M]$. For the compression of the error compensated gradient vector ${\boldsymbol{g}}_{m} \left( \boldsymbol{\theta}_t \right) + \boldsymbol{\Delta}_{m} (t-1)$, we employ the scheme in \cite{DCSattlerSparseBinary}, where it is first sparsified by setting all but the highest $q (t)$ positive and the smallest $q (t)$ negative entries to zero, where $q(t) \le d/2$ (in practice, the goal is to have $q(t) \ll  d$, $\forall t$). Then, device $m$ computes the mean value of the positive and negative entries of the resultant sparse vector, denoted by ${q}_{m}^+(t)$ and ${q}^-_{m}(t)$, respectively, $m \in [M]$. If ${q}_{m}^+(t) \ge {q}_{m}^-(t)$, device $m$ sets all the negative entries of the sparse vector to zero and all the positive entries to ${q}_{m}^+(t)$, and vice versa, if ${q}_{m}^+(t) < {q}_{m}^-(t)$, $m \in [M]$. Let $\hat{\boldsymbol{g}}_{m} \left( \boldsymbol{\theta}_t \right)$ denote the resultant sparse vector at device $m$, $m \in [M]$. After computing $\hat{\boldsymbol{g}}_{m} \left( \boldsymbol{\theta}_t \right)$ at device $m$, $m \in [M]$, the error accumulation vector, which maintains those entries of vector $\boldsymbol{g}_{m} \left(\boldsymbol{\theta}_t \right) + {\boldsymbol{\Delta}_{m} (t-1)}$ that are not transmitted, is updated as follows:  
\begin{align}\label{ErrorAccDDSGD}
\boldsymbol{\Delta}_{m} (t) =
\begin{cases} 
\boldsymbol{g}_{m} \left(\boldsymbol{\theta}_t \right) + {\boldsymbol{\Delta}_{m} (t-1)} - \hat{\boldsymbol{g}}_{m} \left( \boldsymbol{\theta}_t \right), & \mbox{if $m = m^* (t)$},\\
\boldsymbol{g}_{m} \left(\boldsymbol{\theta}_t \right) + {\boldsymbol{\Delta}_{m} (t-1)}, &\mbox{otherwise}. 
\end{cases}
\end{align}
We note that, if user $m$ is scheduled, the accumulated error at device $m$ is the difference between $\boldsymbol{g}_{m} \left(\boldsymbol{\theta}_t \right) + {\boldsymbol{\Delta}_{m} (t-1)}$ and its sparsified version $\hat{\boldsymbol{g}}_{m} \left( \boldsymbol{\theta}_t \right)$, $m \in [M]$; on the other hand, if device $m$ is not scheduled, we maintain vector $\boldsymbol{g}_{m} \left(\boldsymbol{\theta}_t \right) + {\boldsymbol{\Delta}_{m} (t-1)}$ entirely as the accumulated error. For a sparsity level $q (t)$, the D-DSGD scheme requires transmission of a total of \cite[Equation (10)]{MohammadDenizDSGDCS} 
\begin{align}\label{SWMSNumberBits}
r (t) = \log_2 \binom{d}{q (t)} + 33 \mbox{ bits}.
\end{align}
We assume that device $m^* (t)$ employs a capacity achieving channel code using the optimal value of the capacity upper bound in \eqref{CapacityDDSGDRt}, and we set the sparsity level $q (t)$ as the highest integer satisfying $r (t) \le R (t)$.

We highlight here that with the proposed D-DSGD algorithm, all the devices compute the gradient estimates based on the parameter vector received from the PS and their local datasets; however, only a single device is scheduled for transmission over the MAC with the scheduling policy given in \eqref{WorkSelectionDig}. The PS updates the parameter vector after receiving the gradient estimate from the scheduled device and shares it with all the devices to continue their computations.


\begin{remark}\label{RemSelectionCriteria}
An alternative device selection criteria, rather than the one in \eqref{WorkSelectionDig}, is to select the device with the highest capacity upperbound. Here we do not employ this selection criteria due to the overhead introduced by solving the waterfilling power allocation for all $M$ devices. This will be prohibitive for large $s$ and $M$ avlues.
\end{remark}

\begin{remark}\label{Rem2}
Instead of scheduling a single device at each iteration, we can schedule all or a subset of the devices at each iteration, and allocate distinct subchannels to different devices. In Section \ref{SecExperiments} we consider the so-called orthogonal digital DSGD (OD-DSGD) scheme, which schedules all the devices at each iteration, where each device is allocated $\left\lfloor {s/M} \right\rfloor$ distinct subchannels. We have observed that OD-DSGD performs much worse than D-DSGD. It is worth noting that scheduling multiple devices reduces the number of subchannels allocated to each device for orthogonal transmission, and forces the devices to transmit their information at shorter blocklengths. In practice, this would result in a higher error probability or reduced transmission rate  \cite{YoriVinceVerduFiniteBlock}. An alternative approach is to code across time slots by allocating multiple time slots to a scheduled user. This requires the information about the future channel gains, which is not possible in our model since the channel gains are assumed to be i.i.d. across time slots and users.  
\end{remark}

We will evaluate the performance of the D-DSGD scheme in Section \ref{SecExperiments}, and study in detail the impact of various system parameters, such as the average power constraint and the number of devices on the performance. We will also compare the D-DSGD scheme with other compression schemes in the literature, as well as the analog transmission of local gradients, which we present next.  


\section{Analog DSGD}\label{SecPeoposedAnalog}

Analog DSGD is motivated by the fact that the PS is only interested in the average of the gradient vectors, and the underlying wireless MAC can provide the sum of the gradients if they are sent in an uncoded fashion. We first present a generalization of the over-the-air computation approach introduced in \cite{KaibinParallelWork}, referred to as \textit{entry-wise scheduled analog DSGD (ESA-DSGD)}, and then extend it by introducing error accumulation, referred to as \textit{error compensated ESA-DSGD (ECESA-DSGD)}. Finally, we propose a novel analog scheme, built upon our previous work \cite{MohammadDenizDSGDCS}, referred to as \textit{compressed analog DSGD (CA-DSGD)}.  

\subsection{ESA-DSGD}\label{SubSecESscheme}
With the ESA-DSGD scheme studied in \cite{KaibinParallelWork}, each device sends its gradient estimate entirely after applying power allocation, which is to satisfy the average power constraint. At the $t$-th iteration of the DSGD, device $m$, $m \in [M]$, transmits its local gradient estimate ${\boldsymbol{g}}_m \left(\boldsymbol{\theta}_{t} \right) \in \mathbb{R}^{d}$ over $N = \left\lceil {d/2s} \right\rceil$ time slots by utilizing both the real and imaginary components of the available $s$ subchannels. We define, for $n \in [N]$, $m \in [M]$, 
\begin{subequations}
\label{gnmESADSGDDef}
\begin{align}\label{gnmRealESADSGDDef}
{\boldsymbol{g}}^n_{m, {\rm{re}}} \left(\boldsymbol{\theta}_{t} \right)& \triangleq [ g_{m,2(n-1)s+1} \left(\boldsymbol{\theta}_{t} \right), \cdots, g_{m,(2n-1)s} \left(\boldsymbol{\theta}_{t} \right)]^T,\\
{\boldsymbol{g}}^n_{m, {\rm{im}}} \left(\boldsymbol{\theta}_{t} \right)& \triangleq [ g_{m,(2n-1)s+1} \left(\boldsymbol{\theta}_{t} \right), \cdots, g_{m,2ns} \left(\boldsymbol{\theta}_{t} \right)]^T,
\label{gnmImagESADSGDDef}\\
{\boldsymbol{g}}^n_{m} \left(\boldsymbol{\theta}_{t} \right)& \triangleq {\boldsymbol{g}}^n_{m, {\rm{re}}} \left(\boldsymbol{\theta}_{t} \right) + j {\boldsymbol{g}}^n_{m, {\rm{im}}} \left(\boldsymbol{\theta}_{t} \right),
\label{gnmRealImagESADSGDDef}
\end{align}
\end{subequations}
where $g_{m,i} \left(\boldsymbol{\theta}_{t} \right)$ is the $i$-th entry of ${\boldsymbol{g}}_m \left(\boldsymbol{\theta}_{t} \right)$, and we zero-pad ${\boldsymbol{g}}_m \left(\boldsymbol{\theta}_{t} \right)$ to have dimension $2sN$. We note that, according to \eqref{gnmESADSGDDef},
\begin{align}
{\boldsymbol{g}}_m \left(\boldsymbol{\theta}_{t} \right) = [{\boldsymbol{g}}^1_{m, {\rm{re}}} \left(\boldsymbol{\theta}_{t} \right), {\boldsymbol{g}}^1_{m, {\rm{im}}} \left(\boldsymbol{\theta}_{t} \right), \cdots, {\boldsymbol{g}}^N_{m, {\rm{re}}} \left(\boldsymbol{\theta}_{t} \right), {\boldsymbol{g}}^N_{m, {\rm{im}}} \left(\boldsymbol{\theta}_{t} \right)]^T,
\end{align}
where $N = \left\lceil {d/2s} \right\rceil$. At the $n$-th time slot of the $t$-th iteration, device $m$, $m \in [M]$, sends $\boldsymbol{x}^n_{m} \left( t \right) = \boldsymbol{\alpha}^{{\rm{e}}, n}_m (t) \circ {\boldsymbol{g}}^n_m \left(\boldsymbol{\theta}_{t} \right)$, where $\boldsymbol{\alpha}^{{\rm{e}}, n}_m (t) \in \mathbb{C}^{s}$ is the power allocation vector, which is set to satisfy the average transmit power constraint. Thus, after $N$ time slots, each device sends its gradient estimate of dimension $d$ entirely. The $i$-th entry of the power allocation vector $\boldsymbol{\alpha}^{{\rm{e}}, n}_m (t)$ is set as follows: 
\begin{align}\label{ithEntryBetaaVec}
\alpha_{m, i}^{{\rm{e}}, n} (t) =
\begin{cases} 
\frac{\gamma^{{\rm{e}}, n}_m (t)}{h^n_{m,i}(t)}, & \mbox{if $\left| h^n_{m,i}(t) \right|^2 \ge \lambda^{{\rm{e}}} (t)$},\\
0, &\mbox{otherwise}, 
\end{cases}
\end{align}
for some $\gamma^{{\rm{e}}, n}_m (t), \lambda^{{\rm{e}}} (t) \in \mathbb{R}$, set to satisfy the average transmit power constraint. According to \eqref{ithEntryBetaaVec}, each entry of a gradient vector is transmitted if its corresponding channel gain is over a threshold. The set of devices selected to transmit the $i$-th entry of the channel input vector at the $n$-th time slot is given by, $i \in [s]$, $n \in [N]$, 
\begin{align}\label{CalMTheirScheme}
\mathcal{M}^n_i (t) = \left\{ m \in [M]: \left| h_{m,i}^n(t) \right|^2 \ge \lambda^{{\rm{e}}} (t) \right\}.
\end{align}

In the following, we analyze the average transmit power of the ESA-DSGD scheme based on the power allocation design given in \eqref{ithEntryBetaaVec}. 
We set the parameters $\gamma^{{\rm{e}}, n}_m (t)$ and $\lambda^{{\rm{e}}} (t)$ to obtain the same average transmit power $\bar{P}^{n} (t)$ at device $m$, $m \in [M]$, in time slot $n$, $n \in \left[ N \right]$, of iteration $t$, which satisfies
\begin{align}\label{AveragePowerdevicemTheirSchemePbar}
\frac{1}{NT} \sum\nolimits_{t=1}^{T} \sum\nolimits_{n=1}^{N} \bar{P}^{n} (t) \le \bar{P}. 
\end{align}
According to \eqref{ithEntryBetaaVec}, $\forall m \in [M]$, we have
\begin{align}\label{AveragePowerdevicemTheirScheme1}
\bar{P}^{n} (t) & = \mathbb{E} \left[ ||\boldsymbol{x}^n_{m} (t)||^2_2 \right] = \sum\nolimits_{i=1}^{s} \mathbb{E} \left[ \left| \alpha_{m, i}^{{\rm{e}}, n} (t) \right|^2 \left| {g}_{m,(n-1)s+i}(\boldsymbol{\theta}_t) \right|^2 \right], \quad n \in [N], t \in [T].
\end{align}
We highlight that the entries of the gradient vector $\boldsymbol{g}^n_m(\boldsymbol{\theta}_t)$ are independent of the channel gains $h^n_{m,i} (t)$, $\forall i, n, m$. Since the power allocation vector $\boldsymbol{\alpha}^{{\rm{e}}, n}_m (t)$ is a function of $\boldsymbol{g}^n_m(\boldsymbol{\theta}_t)$, it follows that, for $n \in \left[ \left\lceil {d/2s} \right\rceil \right]$, $m \in [M]$,
\begin{align}\label{AveragePowerdevicemTheirScheme2}
& \bar{P}^{n} (t) = \sum\nolimits_{i=1}^{s} \left| {g}_{m,(n-1)s+i}(\boldsymbol{\theta}_t) \right|^2 \mathbb{E} \left[ \left| \alpha_{m, i}^{{\rm{e}}, n} (t) \right|^2 \right].
\end{align}
Note that $\left| h^n_{m,i} (t) \right|^2$ follows an exponential distribution with mean $\sigma^2$, $\forall i, n, m$. Thus, we have
\begin{align}\label{AveragePowerdevicemTheirScheme3}
\mathbb{E} \left[ \left| \alpha_{m, i}^{{\rm{e}}, n} (t) \right|^2 \right] = \left( \frac{ \gamma^{{\rm{e}}, n}_m (t)}{\sigma} \right)^2 {\rm{E}}_1(\lambda^{{\rm{e}}} (t)),
\end{align}
where ${\rm{E}}_1(x) \triangleq \int_x^\infty  \frac{e^{ - \tau }}{\tau} d\tau$. It follows that, $m \in [M], n \in \left[ N \right]$, 
\begin{align}\label{AveragePowerdevicemTheirScheme4}
\bar{P}^{n} (t) = \left( \frac{ \gamma^{{\rm{e}}, n}_m (t)}{\sigma} \right)^2 {\rm{E}}_1(\lambda^{{\rm{e}}}(t)) {P}_{m}^{{\rm{e}}, n} (t), 
\end{align}
where we define ${P}_{m}^{{\rm{e}}, n} (t) \triangleq \left\| {\boldsymbol{g}}^n_m \left(\boldsymbol{\theta}_{t} \right) \right\|_2^2$. Given the threshold value $\lambda^{{\rm{e}}}(t)$, we set
\begin{align}\label{GammaESA}
\gamma^{{\rm{e}}, n}_m (t) = \sigma \bigg( \frac{\bar{P}^{n} (t)}{{\rm{E}}_1(\lambda^{{\rm{e}}}(t)) {P}_{m}^{{\rm{e}}, n} (t)} \bigg)^{1/2}, \quad m \in [M], n \in [N], t \in [T],     
\end{align}
which we note that it does not differ significantly across devices, since values of ${P}_{m}^{{\rm{e}}, n} (t)$, $\forall m$, are not too different. 
We assume that, before transmitting $\boldsymbol{x}^n_{m} \left( t \right) = \boldsymbol{\alpha}^{{\rm{e}}, n}_m (t) \circ {\boldsymbol{g}}^n_m \left(\boldsymbol{\theta}_{t} \right)$, device $m$, $m \in [M]$, sends $\gamma^{{\rm{e}}, n}_m (t)$ to the PS in an error-free fashion using an error correcting code, and the PS computes
\begin{align}
\bar{\gamma}^{{\rm{e}}, n} (t) \triangleq \frac{1}{M} \sum\nolimits_{m=1}^{M} \gamma^{{\rm{e}}, n}_m (t), \quad n \in [N], t \in [T].   
\end{align} 
This factor will be used at the PS to scale down the received signal.

Here we analyze the received signal at the PS. By substituting $\boldsymbol{x}^n_{m} \left(\boldsymbol{\theta}_{t} \right)$ and $\boldsymbol{\alpha}^{{\rm{e}}, n}_m (t)$ into \eqref{ReceivedVectorPSGen}, it follows that, for $i \in [s]$, $n \in [N]$,
\begin{align}\label{ReceivedVectorPSGenTheirScheme}
{y}_i^n (t) =  \sum\nolimits_{m \in \mathcal{M}_i^n (t)} \gamma^{{\rm{e}},n}_m (t) \left( g_{m, 2(n-1)s+i} (\boldsymbol{\theta}_t) + j g_{m, (2n-1)s+i} (\boldsymbol{\theta}_t) \right) + {z}_i^n (t).  
\end{align}
The PS has perfect CSI, and hence, knows set $\mathcal{M}_i^n (t)$. Its goal is to recover $\frac{1}{\left| \mathcal{M}_i^n (t) \right|}\sum\nolimits_{m \in \mathcal{M}_i^n (t)} {g}_{m,2(n-1)s+i} \left(\boldsymbol{\theta}_{t} \right)$ and $\frac{1}{\left| \mathcal{M}_i^n (t) \right|}\sum\nolimits_{m \in \mathcal{M}_i^n (t)} {g}_{m,(2n-1)s+i} \left(\boldsymbol{\theta}_{t} \right)$, which provide estimates for $\frac{1}{M} \sum\nolimits_{m=1}^{M} {g}_{m,2(n-1)s+i} \left(\boldsymbol{\theta}_{t} \right)$ and $\frac{1}{M} \sum\nolimits_{m=1}^{M} {g}_{m,(2n-1)s+i} \left(\boldsymbol{\theta}_{t} \right)$, respectively. The PS estimates $\frac{1}{\left| \mathcal{M}_i^n (t) \right|}\sum\nolimits_{m \in \mathcal{M}_i^n (t)} {g}_{m,2(n-1)s+i} \left(\boldsymbol{\theta}_{t} \right)$, for $i \in [s]$, $n \in [N]$, using its noisy observation ${y}_i^n(t)$, given in \eqref{ReceivedVectorPSGenTheirScheme}, as 
\begin{align}\label{EstimateToUpdateTheirScheme}
\hat{{g}}^{\rm{e}}_{2(n-1)s+i} \left(\boldsymbol{\theta}_{t} \right) = 
\begin{cases} 
\frac{{\rm{Re}} \{ {y}_i^n(t) \}}{\bar{\gamma}^{{\rm{e}}, n} (t) \left| \mathcal{M}_i^n (t) \right|}, & \mbox{if $\left| \mathcal{M}_i^n (t) \right| \ne 0$},\\
0, &\mbox{otherwise}, 
\end{cases}
\end{align}
and estimates $\frac{1}{\left| \mathcal{M}_i^n (t) \right|}\sum\nolimits_{m \in \mathcal{M}_i^n (t)} {g}_{m,(2n-1)s+i} \left(\boldsymbol{\theta}_{t} \right)$ through 
\begin{align}\label{ImagEstimateToUpdateTheirScheme}
\hat{{g}}^{\rm{e}}_{(2n-1)s+i} \left(\boldsymbol{\theta}_{t} \right) = 
\begin{cases} 
\frac{{\rm{Im}} \{ {y}_i^n(t) \}}{\bar{\gamma}^{{\rm{e}}, n} \left| \mathcal{M}_i^n (t) \right|}, & \mbox{if $\left| \mathcal{M}_i^n (t) \right| \ne 0$},\\
0, &\mbox{otherwise}. 
\end{cases}
\end{align}
Estimated vector $\hat{\boldsymbol{g}}^{\rm{e}} \left(\boldsymbol{\theta}_{t} \right) \triangleq [ \hat{g}^{\rm{e}}_{1} \left(\boldsymbol{\theta}_{t} \right) \cdots \hat{g}^{\rm{e}}_{d} \left(\boldsymbol{\theta}_{t} \right)]^T$ is then used to update the parameter vector as $\boldsymbol{\theta}_{t+1} =\boldsymbol{\theta}_{t} - \eta_t \hat{\boldsymbol{g}}^{\rm{e}} \left(\boldsymbol{\theta}_{t} \right)$.

\begin{remark}\label{RemKaibinInsPower}
We remark here that the scheme in \cite{KaibinParallelWork} imposes a stricter average power constraint $\bar{P}$ per iteration of the DSGD, i.e., at device $m$ we should have
\begin{align}\label{AvePowerConsKaibinsPaper}
\frac{1}{N} \sum\nolimits_{n=1}^{N} \mathbb{E} \left[ ||\boldsymbol{x}^n_{m} (t)||^2_2 \right] \le \bar{P}, \quad \forall m \in [M], \forall t. 
\end{align}
For fairness in our comparisons we relax this power constraint, and impose the one in \eqref{AvePowerConsGen}, which constrains the average power over all the iterations.  
\end{remark}

\subsection{ECESA-DSGD}\label{SubSecECESAscheme}
With the ESA-DSGD scheme, entries of the gradient vectors that are not sent due to poor channel conditions are completely forgotten. The proposed ECESA-DSGD scheme modifies ESA-DSGD by incorporating error accumulation technique to retain the accuracy of local gradients.

We denote the error accumulation vector calculated by device $m$ at the $n$-th time slot of the $t$-iteration by ${\boldsymbol{\Delta}_{m}^{{\rm{v}}, n} (t)} \in \mathbb{C}^s$, and set ${\boldsymbol{\Delta}_{m}^{{\rm{v}}, n} (t)} = \boldsymbol{0}$, $\forall n,t,m$. Similarly to the ESA-DSGD scheme, with ECESA-DSGD, each device sends its entire gradient estimate of dimension $d$ through $N = \left\lceil {d/2s} \right\rceil$ time slots, where the gradient estimates at the devices are zero-padded to dimension $2sN$. After computing $\boldsymbol{g}_m \left(\boldsymbol{\theta}_t \right)$ and obtaining $\boldsymbol{g}_m^{n} \left(\boldsymbol{\theta}_t \right)$ according to \eqref{gnmESADSGDDef}, device $m$, $m \in [M]$, updates its gradient estimate with the accumulated error as $\boldsymbol{g}_m^{{\rm{v}}, n} \left(\boldsymbol{\theta}_t \right) \triangleq \boldsymbol{g}_m^{n} \left(\boldsymbol{\theta}_t \right) + {\boldsymbol{\Delta}_{m}^{{\rm{v}}, n} (t - 1)}$, for $n \in [N]$, and transmits vector $\boldsymbol{x}^n_{m} \left( t \right) = \boldsymbol{\alpha}^{{\rm{v}}, n}_m (t) \circ {\boldsymbol{g}}^{{\rm{v}}, n}_m \left(\boldsymbol{\theta}_{t} \right)$, where $\boldsymbol{\alpha}^{{\rm{v}}, n}_m (t) \in \mathbb{C}^{s}$ is the power allocation vector, whose $i$-th entry is determined as follows:
\begin{align}\label{ithEntryECESAaVec}
\alpha_{m, i}^{{\rm{v}}, n} (t) =
\begin{cases} 
\frac{\gamma^{{\rm{v}}, n}_m (t)}{h^n_{m,i}(t)}, & \mbox{if $\left| h^n_{m,i}(t) \right|^2 \ge \lambda^{{\rm{v}}} (t)$},\\
0, &\mbox{otherwise}, 
\end{cases}
\end{align}
for some $\gamma^{{\rm{v}}, n}_m (t), \lambda^{{\rm{v}}} (t) \in \mathbb{R}$. Device $m$, $m \in [M]$, then updates the $i$-th entry of vector ${\boldsymbol{\Delta}_{m}^{{\rm{v}}, n} (t)}$ as follows:
\begin{align}\label{DeltaECESA}
{{\Delta}_{m, i}^{{\rm{v}}, n} (t)} &= \left( 1 - \mathds{1} \left( \alpha_{m, i}^{{\rm{v}}, n} (t) \ne 0 \right) \right) {{g}}_{m,i}^{n} \left(\boldsymbol{\theta}_{t} \right) \nonumber\\
& = \left( 1 - \mathds{1} \left( \alpha_{m, i}^{{\rm{v}}, n} (t) \ne 0 \right) \right) \left( {{g}}_{m,2(n-1)s+i} \left(\boldsymbol{\theta}_{t} \right) + j {{g}}_{m,(2n-1)s+i} \left(\boldsymbol{\theta}_{t} \right) \right),  
\end{align}
where $\mathds{1} (\cdot)$ is the indicator function, and ${{g}}_{m,i}^{n} \left(\boldsymbol{\theta}_{t} \right)$ denotes the $i$-th entry of $\boldsymbol{g}_m^{n} \left(\boldsymbol{\theta}_t \right)$, for $i \in [s]$, $n \in [N]$. Thus, the $i$-th entry of vector ${\boldsymbol{g}}^{{\rm{v}}, n}_m \left(\boldsymbol{\theta}_{t} \right)$ is given by, for $i \in [s]$, $n \in [N]$,
\begin{align}\label{ithEntryVecgvn}
{{g}}^{{\rm{v}}, n}_{m,i} \left(\boldsymbol{\theta}_{t} \right) = & {{g}}_{m,2(n-1)s+i} \left(\boldsymbol{\theta}_{t} \right) + j {{g}}_{m,(2n-1)s+i} \left(\boldsymbol{\theta}_{t} \right) \nonumber\\
& + \left( 1 - \mathds{1} \left( \alpha_{m, i}^{{\rm{v}}, n} (t-1) \ne 0 \right) \right) \left( {{g}}_{m,2(n-1)s+i} \left(\boldsymbol{\theta}_{t-1} \right) + j {{g}}_{m,(2n-1)s+i} \left(\boldsymbol{\theta}_{t-1} \right) \right).
\end{align}
According to \eqref{DeltaECESA}, each entry of the gradient vector ${\boldsymbol{g}}^{{\rm{v}}, n}_m \left(\boldsymbol{\theta}_{t} \right)$ that is not transmitted due to the power allocation given in \eqref{ithEntryECESAaVec}, is retained in the error accumulation vector ${\boldsymbol{\Delta}_{m}^{{\rm{v}}, n} (t)}$ for possible transmission in the next iteration.

Here we provide the power analysis of the ECESA-DSGD scheme. For fairness, we set the parameters $\gamma^{{\rm{v}}, n}_m (t)$ and $\lambda^{{\rm{v}}} (t)$ yielding an average transmit power $\bar{P}^{n} (t)$ at device $m$, $m \in [M]$, in time slot $n$, $n \in \left[ N \right]$, of iteration $t$, satisfying the constraint in \eqref{AveragePowerdevicemTheirSchemePbar}. Since the power allocation vector of ECESA-DSGD, given in \eqref{ithEntryECESAaVec}, is similar to that of the ESA-DSGD, by following a similar procedure we obtain the following average power at device $m$ for ECESA-DSGD:
\begin{align}\label{AveragePowerdevicemTheirSchemeECESA}
\bar{P}^{n} (t) = \left( \frac{ \gamma^{{\rm{v}}, n}_m (t)}{\sigma} \right)^2 {\rm{E}}_1(\lambda^{{\rm{v}}} (t)) {P}_{m}^{{\rm{v}}, n} (t), \quad n \in [N], t \in [T], 
\end{align} 
where we define ${P}_{m}^{{\rm{v}}, n} (t) \triangleq \left\| {\boldsymbol{g}}^{{\rm{v}}, n}_m \left(\boldsymbol{\theta}_{t} \right) \right\|_2^2$. For a fixed $\lambda^{{\rm{v}}}(t)$, we set
\begin{align}\label{GammaECESA}
\gamma^{{\rm{v}}, n}_m (t) = \sigma \bigg( \frac{\bar{P}^{n} (t)}{{\rm{E}}_1(\lambda^{{\rm{e}}}(t)) {P}_{m}^{{\rm{v}}, n} (t)} \bigg)^{1/2}, \quad m \in [M], n \in [N], t \in [T],     
\end{align}
shared with the PS in an error-free manner, through which the PS computes
\begin{align}
\bar{\gamma}^{{\rm{v}}, n} (t) \triangleq \frac{1}{M} \sum\nolimits_{m=1}^{M} \gamma^{{\rm{v}}, n}_m (t), \quad n \in [N], t \in [T].   
\end{align}

From the power allocation in \eqref{ithEntryECESAaVec}, it follows that, for $i \in [s]$, $n \in [N]$,
\begin{align}\label{ReceivedVectorPSGenTheirSchemeECESA}
{y}_i^n (t) =  \sum\nolimits_{m \in \mathcal{M}_i^n (t)} \gamma^{{\rm{v}}, n}_m (t) g_{m, i}^{{\rm{v}}, n} (\boldsymbol{\theta}_t)  + {z}_i^n (t),  
\end{align} 
where we have 
\begin{align}\label{CalMTheirSchemeECESA}
\mathcal{M}^n_i (t) = \left\{ m \in [M]: \left| h_{m,i}^n(t) \right|^2 \ge \lambda^{{\rm{v}}} (t) \right\}.
\end{align}
Having perfect CSI, the PS's goal is to recover $\frac{1}{\left| \mathcal{M}_i^n (t) \right|}\sum\nolimits_{m \in \mathcal{M}_i^n (t)} {g}_{m,i}^{{\rm{v}},n} \left(\boldsymbol{\theta}_{t} \right)$, the real and imaginary terms of which provide estimates for $\frac{1}{M} \sum\nolimits_{m=1}^{M} {g}_{m,2(n-1)s+i}^{n} \left(\boldsymbol{\theta}_{t} \right)$ and $\frac{1}{M} \sum\nolimits_{m=1}^{M} {g}_{m,(2n-1)s+i}^{n} \left(\boldsymbol{\theta}_{t} \right)$, respectively, for $i \in [s]$, $n \in [N]$. The PS estimates $\frac{1}{M} \sum\nolimits_{m=1}^{M} {g}_{m,2(n-1)s+i}^{n} \left(\boldsymbol{\theta}_{t} \right)$ as  
\begin{align}\label{EstimateToUpdateTheirSchemeECESAReal}
\hat{{g}}^{{\rm{v}}}_{2(n-1)s+i} \left(\boldsymbol{\theta}_{t} \right) = 
\begin{cases} 
\frac{{\rm{Re}} \{ {y}_i^n(t) \}}{\bar{\gamma}^{{\rm{v}}, n} (t) \left| \mathcal{M}_i^n (t) \right|}, & \mbox{if $\left| \mathcal{M}_i^n (t) \right| \ne 0$},\\
\hat{{g}}^{{\rm{v}}}_{2(n-1)s+i} \left(\boldsymbol{\theta}_{t-1} \right), &\mbox{otherwise}, 
\end{cases}
\end{align}
and estimates $\frac{1}{M} \sum\nolimits_{m=1}^{M} {g}_{m,(2n-1)s+i}^{n} \left(\boldsymbol{\theta}_{t} \right)$ through
\begin{align}\label{EstimateToUpdateTheirSchemeECESAImag}
\hat{{g}}^{{\rm{v}}}_{(2n-1)s+i} \left(\boldsymbol{\theta}_{t} \right) = 
\begin{cases} 
\frac{{\rm{Im}} \{ {y}_i^n(t) \}}{\bar{\gamma}^{{\rm{v}}, n} (t) \left| \mathcal{M}_i^n (t) \right|}, & \mbox{if $\left| \mathcal{M}_i^n (t) \right| \ne 0$},\\
\hat{{g}}^{{\rm{v}}}_{(2n-1)s+i} \left(\boldsymbol{\theta}_{t-1} \right), &\mbox{otherwise}, 
\end{cases}
\end{align}
for $i \in [s]$, $n \in [N]$. After $N = \left\lceil {d/2s} \right\rceil$ time slots, estimated vector $\hat{\boldsymbol{g}}^{\rm{v}} \left(\boldsymbol{\theta}_{t} \right) \triangleq [ \hat{{g}}^{{\rm{v}}}_{1}, \cdots, \hat{{g}}^{{\rm{v}}}_{d}]^T$ is then used to update the parameter vector as $\boldsymbol{\theta}_{t+1} =\boldsymbol{\theta}_{t} - \eta_t \hat{\boldsymbol{g}}^{\rm{v}} \left(\boldsymbol{\theta}_{t} \right)$.

\subsection{CA-DSGD}\label{SubSecCWSscheme}
As opposed to ESA-DSGD and ECESA-DSGD, which aim to transmit all the gradient entries to the PS at each DSGD iteration, i.e., $N = \left\lceil {d/2s} \right\rceil$, the CA-DSGD scheme proposed here reduces the transmission bandwidth by reducing the dimension of the gradient vector by a linear projection. Each device projects its gradient estimate to dimension $\tilde{s} = 2 s N$, which can then be transmitted through $N$ time slots, for some $N \in \left[ \left\lceil {d/2s} \right\rceil \right]$. 
The details of CA-DSGD are given in Algorithm \ref{A-DSGD_alg}.

We describe the CA-DSGD scheme for an arbitrary number of time slots $N \in \left[ \left\lceil {d/2s} \right\rceil \right]$ per iteration of DSGD, which is determined later. At each iteration the devices sparsify their gradient estimates as described below. They employ \textit{error accumulation} \cite{DCOneBitQuan}, where the accumulated error vector at device $m$ until iteration $t$ is denoted by ${\boldsymbol{\Delta}_{m}^{{\rm{c}}} (t - 1)} \in \mathbb{R}^{d}$, where we set ${\boldsymbol{\Delta}^{{\rm{c}}}_{m}(0)} = \boldsymbol{0}$, $\forall m \in [M]$. After computing $\boldsymbol{g}_m \left(\boldsymbol{\theta}_t \right)$, device $m$ updates its estimate with the accumulated error as $\boldsymbol{g}_m^{ec} \left(\boldsymbol{\theta}_t \right) \triangleq \boldsymbol{g}_m \left(\boldsymbol{\theta}_t \right) + {\boldsymbol{\Delta}^{{\rm{c}}}_{m} (t - 1)}$, $m \in [M]$. Next, the devices apply gradient sparsification, where device $m$ sets all but $k$ elements with the highest magnitudes of vector $\boldsymbol{g}_m^{ec} \left(\boldsymbol{\theta}_t \right)$ to zero, where $k \le \tilde{s}$ is a design parameter, and obtains a sparse vector $\boldsymbol{g}_m^{sp} \left(\boldsymbol{\theta}_t \right)$, $m \in [M]$. This $k$-level sparsification is represented by function ${\rm{sparse}}_k$ in Algorithm \ref{ModelUpdateAlg}, i.e., $\boldsymbol{g}_m^{sp} \left(\boldsymbol{\theta}_t \right) = {\rm{sparse}}_k \left( \boldsymbol{g}_m^{ec} \left(\boldsymbol{\theta}_t \right) \right)$. device $m$, $m \in [M]$, then updates $\boldsymbol{\Delta}_{m}^{{\rm{c}}} (t)$ as $\boldsymbol{\Delta}^{{\rm{c}}}_{m} (t) = \boldsymbol{g}_m^{ec} \left(\boldsymbol{\theta}_t \right) - \boldsymbol{g}_m^{sp} \left(\boldsymbol{\theta}_t \right)$. To transmit the sparse vectors over the limited-bandwidth channel, devices employ a random projection matrix, similarly to compressive sensing.

\begin{algorithm}[t]
\caption{CA-DSGD}
\label{ModelUpdateAlg}
\begin{algorithmic}[1]
\Statex
\State{\textbf{Initialize} $\boldsymbol{\theta}_1 = \boldsymbol{0}$ and $\boldsymbol{\Delta}^{\rm{c}}_{1} (0) = \cdots = \boldsymbol{\Delta}^{\rm{c}}_{M} (0) = \boldsymbol{0}$}
\For {$t = 1, \ldots, T$}
\Statex
\begin{itemize}
\item \textbf{devices do:}
\end{itemize}
\For {$m = 1, \ldots, M$ in parallel}
\State{Compute $\boldsymbol{g}_m \left( \boldsymbol{\theta}_t \right)$ with respect to local dataset $\mathcal{B}_{m}$}
\State{$\boldsymbol{g}_m^{ec} \left( \boldsymbol{\theta}_t \right) = \boldsymbol{g}_m \left( \boldsymbol{\theta}_t \right) + {\boldsymbol{\Delta}^{\rm{c}}_{m} ( t - 1)}$}
\State{$\boldsymbol{g}_m^{sp} \left( \boldsymbol{\theta}_t \right) = {\rm{sparse}}_k \left( \boldsymbol{g}_m^{ec} \left( \boldsymbol{\theta}_t \right) \right)$}
\State{$\boldsymbol{\Delta}_{m}^{\rm{c}} (t) = \boldsymbol{g}_m^{ec} \left( \boldsymbol{\theta}_t \right) - \boldsymbol{g}_m^{sp} \left( \boldsymbol{\theta}_t \right)$}
\State{$\tilde{\boldsymbol{g}}_m \left( \boldsymbol{\theta}_{t} \right) = \boldsymbol{A} \boldsymbol{g}_m^{sp} \left( \boldsymbol{\theta}_t \right)$}
\For {$n = 1, \ldots, N$}
\State{$\boldsymbol{x}^n_{m} \left( t \right) = \boldsymbol{\alpha}^{{\rm{c}}, n}_m (t) \circ \tilde{\boldsymbol{g}}^n_m \left(\boldsymbol{\theta}_{t} \right)$}
\EndFor
\EndFor
\Statex
\begin{itemize}
\item \textbf{PS does:}
\end{itemize}
\If {$\hat{\boldsymbol{y}} (t) \ne \boldsymbol{0}$}
    \State $\hat{\boldsymbol{g}}^{\rm{c}} \left(\boldsymbol{\theta}_{t} \right) = {\rm{AMP}}_{\boldsymbol{A}} \left( \hat{\boldsymbol{y}} (t) \right) $
    \State $\boldsymbol{\theta}_{t+1} =\boldsymbol{\theta}_{t} - \eta_t \hat{\boldsymbol{g}}^{\rm{c}} \left(\boldsymbol{\theta}_{t} \right)$
\Else
    \State $\boldsymbol{\theta}_{t+1} =\boldsymbol{\theta}_{t}$
\EndIf
\EndFor
\end{algorithmic}\label{A-DSGD_alg}
\end{algorithm}

A pseudo-random matrix $\boldsymbol{A} \in \mathbb{R}^{{\tilde{s}} \times d}$, with each entry i.i.d. according to $\mathcal{N} (0,1/{\tilde{s}})$, is generated and shared between the PS and the devices, where $\tilde{s} = 2 s N$, for an arbitrary $N \in \left[ \left\lceil {d/2s} \right\rceil \right]$. At each iteration $t$, device $m$ computes $\tilde{\boldsymbol{g}}_m \left(\boldsymbol{\theta}_{t} \right) \triangleq \boldsymbol{A} \boldsymbol{g}_m^{sp} \left(\boldsymbol{\theta}_t \right) \in \mathbb{R}^{{\tilde{s}}}$, and aims to transmit it to the PS over $N = {\tilde{s}/2s}$ time slots. We define, for $n \in [N]$, $m \in [M]$, 
\begin{subequations}
\label{gnmCADSGDDef}
\begin{align}\label{gnmRealCADSGDDef}
\tilde{\boldsymbol{g}}^n_{m, {\rm{re}}} \left(\boldsymbol{\theta}_{t} \right) &\triangleq [ \tilde{g}_{m,2(n-1)s+1} \left(\boldsymbol{\theta}_{t} \right), \cdots, \tilde{g}_{m,(2n-1)s} \left(\boldsymbol{\theta}_{t} \right)]^T,\\
\tilde{\boldsymbol{g}}^n_{m, {\rm{im}}} \left(\boldsymbol{\theta}_{t} \right) &\triangleq [ \tilde{g}_{m,(2n-1)s+1} \left(\boldsymbol{\theta}_{t} \right), \cdots, \tilde{g}_{m,2ns} \left(\boldsymbol{\theta}_{t} \right)]^T,
\label{gnmImagCADSGDDef}\\
\tilde{\boldsymbol{g}}^n_{m} \left(\boldsymbol{\theta}_{t} \right) &\triangleq \tilde{\boldsymbol{g}}^n_{m, {\rm{re}}} \left(\boldsymbol{\theta}_{t} \right) + j \tilde{\boldsymbol{g}}^n_{m, {\rm{im}}} \left(\boldsymbol{\theta}_{t} \right),
\label{gnmRealImagCADSGDDef}
\end{align}
\end{subequations}
where $\tilde{g}_{m,i} \left(\boldsymbol{\theta}_{t} \right)$ is the $i$-th entry of $\tilde{\boldsymbol{g}}_m \left(\boldsymbol{\theta}_{t} \right)$, $i \in [\tilde{s}]$. At the $n$-th time slot of the $t$-th iteration of DSGD, device $m$, $m \in [M]$, sends $\boldsymbol{x}^n_{m} \left( t \right) = \boldsymbol{\alpha}^{{\rm{c}}, n}_m (t) \circ \tilde{\boldsymbol{g}}^n_m \left(\boldsymbol{\theta}_{t} \right)$, where $\boldsymbol{\alpha}^{{\rm{c}}, n}_m (t) \in \mathbb{C}^{s}$ is the power allocation vector. The $i$-th entry of the power allocation vector $\boldsymbol{\alpha}^{{\rm{c}}, n}_m (t)$ is set as follows: 
\begin{align}\label{ithEntryAlphaVec}
\alpha_{m, i}^{{\rm{c}}, n} (t) =
\begin{cases} 
\frac{\gamma^{{\rm{c}}, n}_m (t)}{h^n_{m,i}(t)}, & \mbox{if $\left| h^n_{m,i}(t) \right|^2 \ge \lambda^{{\rm{c}}} (t)$},\\
0, &\mbox{otherwise}, 
\end{cases}
\end{align}
for some $\gamma^{{\rm{c}}, n}_m (t), \lambda^{{\rm{c}}} (t) \in \mathbb{R}$. The set of devices scheduled to transmit the $i$-th entry of the channel input vector at the $n$-th time slot is given by, $i \in [s]$, $n \in [N]$, 
\begin{align}\label{CalMMyScheme}
\mathcal{M}^n_i (t) = \left\{ m \in [M]: \left| h_{m,i}^n(t) \right|^2 \ge \lambda^{{\rm{c}}, n}_{m} (t) \right\}.
\end{align}

Similarly to ESA-DSGD and ECESA-DSGD, we set the average transmit power at device $m$, $m \in [M]$, in time slot $n$, $n \in \left[ N \right]$, of iteration $t$ for CA-DSGD to $\bar{P}^{n} (t)$,  
\begin{align}\label{AveragePowerdevicemTheirSchemeCA}
\bar{P}^{n} (t) = \left( \frac{ \gamma^{{\rm{c}}, n}_m (t)}{\sigma} \right)^2 {\rm{E}}_1(\lambda^{{\rm{c}}} (t)) {P}_{m}^{{\rm{c}}, n} (t), \quad t \in [T], 
\end{align}
where we define ${P}_{m}^{{\rm{c}}, n} (t) \triangleq \left\| \tilde{\boldsymbol{g}}^{n}_m \left(\boldsymbol{\theta}_{t} \right) \right\|_2^2$. Given $\lambda^{{\rm{c}}}(t)$, we set
\begin{align}\label{GammaCA}
\gamma^{{\rm{c}}, n}_m (t) = \sigma \bigg( \frac{\bar{P}^{n} (t)}{{\rm{E}}_1(\lambda^{{\rm{e}}}(t)) {P}_{m}^{{\rm{c}}, n} (t)} \bigg)^{1/2}, \quad m \in [M], n \in [N], t \in [T],     
\end{align}
and the PS computes 
\begin{align}
\bar{\gamma}^{{\rm{c}}, n} (t) \triangleq \frac{1}{M} \sum\nolimits_{m=1}^{M} \gamma^{{\rm{c}}, n}_m (t), \quad n \in [N], t \in [T],   
\end{align}
after receiving $\gamma^{{\rm{c}}, n}_m (t)$.

By substituting $\boldsymbol{x}^n_{m} \left(\boldsymbol{\theta}_{t} \right)$ and $\boldsymbol{\alpha}^{{\rm{c}}} (t)$ into \eqref{ReceivedVectorPSGen}, it follows that, for $i \in [s]$, $n \in [N]$
\begin{align}\label{ReceivedVectorPSGenOurScheme}
& {y}_i^n (t) = \sum\nolimits_{m \in \mathcal{M}_i^n (t)} \gamma^{{\rm{c}}, n}_m (t) \left( \tilde{g}_{m, 2(n-1)s+i} (\boldsymbol{\theta}_t) + j \tilde{g}_{m, (2n-1)s+i} (\boldsymbol{\theta}_t) \right) + {z}_i^n (t) \nonumber\\
& \;\;\; = \boldsymbol{a}_{2(n-1)s+i}^T \sum\nolimits_{m \in \mathcal{M}_i^n (t)} \gamma^{{\rm{c}}, n}_m (t) {\boldsymbol{g}}^{sp}_{m} (\boldsymbol{\theta}_t) + j \boldsymbol{a}_{(2n-1)s+i}^T \sum\nolimits_{m \in \mathcal{M}_i^n (t)} \gamma^{{\rm{c}}, n}_m (t) {\boldsymbol{g}}^{sp}_{m} (\boldsymbol{\theta}_t) + {z}_i^n (t), 
\end{align}
where $\boldsymbol{a}_{i}^T$ denotes the $i$-th row of measurement matrix $\boldsymbol{A}$, and we note that $\tilde{g}_{m, i} (\boldsymbol{\theta}_t) = \boldsymbol{a}_{i}^T {\boldsymbol{g}}^{sp}_{m} (\boldsymbol{\theta}_t)$, $i \in [\tilde{s}]$. The PS wants to recover $\frac{1}{M}\sum\nolimits_{m=1}^{M} \boldsymbol{g}_m^{sp} \left(\boldsymbol{\theta}_{t} \right)$ from its noisy observations in \eqref{ReceivedVectorPSGenOurScheme}. For this, using its knowledge of matrix $\boldsymbol{A}$ and the CSI, PS employs the approximate message passing (AMP) algorithm \cite{AMPJournalDonoho}. The AMP algorithm is represented by ${\rm{AMP}}_{\boldsymbol{A}}$ in Algorithm \ref{ModelUpdateAlg}. The PS first obtains, for $i \in [s]$, $n \in [N]$,
\begin{subequations}
\label{yhatntCA}
\begin{align}\label{yhatntCAReal}
\hat{y}_{2(n-1)s+i}(t)=
\begin{cases} 
\frac{{\rm{Re}} \{ y_i^n (t) \}}{\bar{\gamma}^{{\rm{c}}, n}_m (t) \left| \mathcal{M}_i^n (t) \right|}, & \mbox{if $\left| \mathcal{M}_i^n (t) \right| \ne 0$},\\
0, &\mbox{otherwise}, 
\end{cases}\\
\hat{y}_{(2n-1)s+i}(t)=
\begin{cases} 
\frac{{\rm{Im}} \{ y_i^n (t) \}}{\bar{\gamma}^{{\rm{c}}, n}_m (t) \left| \mathcal{M}_i^n (t) \right|}, & \mbox{if $\left| \mathcal{M}_i^n (t) \right| \ne 0$},\\
0, &\mbox{otherwise}, 
\end{cases}
\label{yhatntCAImag}
\end{align}
\end{subequations}
and then estimates
\begin{align}\label{EstimateToUpdateOurScheme}
\hat{\boldsymbol{g}}^{\rm{c}} \left(\boldsymbol{\theta}_{t} \right) = {\rm{AMP}}_{\boldsymbol{A}} \left( \hat{\boldsymbol{y}} (t) \right),
\end{align}
where we define $\hat{\boldsymbol{y}} (t) \triangleq [ \hat{{y}}_{1} (t), \cdots, \hat{{y}}_{\tilde{s}} (t)]^T$. If $\hat{\boldsymbol{y}} (t) \ne \boldsymbol{0}$, $\hat{\boldsymbol{g}}^{\rm{c}} \left(\boldsymbol{\theta}_{t} \right)$ is used to update the parameter vector as $\boldsymbol{\theta}_{t+1} =\boldsymbol{\theta}_{t} - \eta_t \hat{\boldsymbol{g}}^{\rm{c}} \left(\boldsymbol{\theta}_{t} \right)$. On the other hand, if $\hat{\boldsymbol{y}} (t) = \boldsymbol{0}$, the previous parameter vector is simply used as the new one, i.e., $\boldsymbol{\theta}_{t+1} =\boldsymbol{\theta}_{t}$.

\begin{remark}\label{CAtoECESA}
For $N = \left\lceil {d/2s} \right\rceil$, in which the entire gradient vectors are transmitted to the PS at each iteration, the CA-DSGD scheme reduces to the ECESA-DSGD scheme; that is, sparsification and projection to a vector of smaller dimension, given in lines 6 and 8 of Algorithm \ref{ModelUpdateAlg}, respectively, are not performed at the devices. 
\end{remark}

\begin{remark}\label{CAk}
We remark that $k$ is a design parameter which can take different values limited to $k < \tilde{s}$. For relatively small $k$ values, $\frac{1}{M} \sum\nolimits_{m=1}^{M} \boldsymbol{g}_m^{sp} \left( \boldsymbol{\theta}_t \right)$ can be recovered from $\frac{1}{M} \sum\nolimits_{m=1}^{M} \tilde{\boldsymbol{g}}_m \left( \boldsymbol{\theta}_t \right)$ in a more reliable manner as we have more noisy measurements of the sparsified gradient estimate than the number of its non-zero elements; however, $\frac{1}{M} \sum\nolimits_{m=1}^{M} \boldsymbol{g}_m^{sp} \left( \boldsymbol{\theta}_t \right)$ provides a less accurate estimate of the actual average gradient $\frac{1}{M} \sum\nolimits_{m=1}^{M} \boldsymbol{g}_m \left( \boldsymbol{\theta}_t \right)$ as we have removed more componenets from the gradient estimates. On the other hand, for a relatively high $k$ value, $\frac{1}{M} \sum\nolimits_{m=1}^{M} \boldsymbol{g}_m^{sp} \left( \boldsymbol{\theta}_t \right)$ provides a better estimate for $\frac{1}{M} \sum\nolimits_{m=1}^{M} \boldsymbol{g}_m \left( \boldsymbol{\theta}_t \right)$; however, it becomes harder to recover $\frac{1}{M} \sum\nolimits_{m=1}^{M} \boldsymbol{g}_m^{sp} \left( \boldsymbol{\theta}_t \right)$ from $\frac{1}{M} \sum\nolimits_{m=1}^{M} \tilde{\boldsymbol{g}}_m \left( \boldsymbol{\theta}_t \right)$ in a reliable manner.     
\end{remark}

\begin{remark}\label{CorrelationGradientEstimatesAtdevices}
We note that, even though each device transmits a sparse vector $\boldsymbol{g}_m^{sp} \left( \boldsymbol{\theta}_t \right)$, their sum received over the channel does not need to be sparse. However, when the datasets are i.i.d. across devices and $B$ is large, we expect the gradient estimates across devices to be statistically uniform, and thus, will have similar sparsity patterns. Note, however, that the proposed CA-DSGD scheme does not require data to be independent across devices. 
As it will be shown in Fig. \ref{Figure_analog_digital_IIDnonIID}, the CA-DSGD scheme converges even when the local datasets across the devices are biased, i.e., non-IID data distribution scenario described later, where the sparsity patterns of the local gradient estimates are expected to be more diverse. We observe that the transmissions from multiple devices still align on a small number of coordinates thanks to the superposition property of analog transmission. Thus, AMP still manages to recover the average gradient with reasonable accuracy, and the DSGD process converges, albeit more slowly compared to the IID data distribution scenario. We will see in Fig. \ref{Figure_analog_digital_IIDnonIID} that CA-DSGD outperforms alternative analog and digital schemes with even a higher performance gap in the non-IID data distribution scenario. 
\end{remark}

\begin{remark}\label{RemDifESACWSA}
With ESA-DSGD, each device transmits only the entries of its estimated gradient whose corresponding channel conditions are sufficiently good. Thus, the gradient vector is inherently sparsified, but only based on the channel gains, regardless of the importance of the gradient entries. Then the entire gradient vector is sent over the bandwidth-limited wireless MAC over orthogonal time periods. On the other hand, with CA-DSGD, each device sends only $k \le \tilde{s}$ important gradient entries, where the magnitude of each entry is regarded as the importance metric, by projecting the sparse gradient vector to a low-dimensional vector of length $\tilde{s} \le d$. We further highlight the error accumulation technique incorporated into ECESA-DSGD and CA-DSGD, whereas with ESA-DSGD, entries of the gradient vectors that are not sent are forgotten.    
\end{remark}

\begin{remark}\label{RemAnalogPrivacy}
We highlight that, thanks to the wireless MAC providing a noisy version of the average of the gradient estimates, the analog schemes can potentially help preserve the privacy as well. This is particularly compelling for the proposed CA-DSGD scheme, where the gradient estimates are compressed through linear projection before transmission.  
\end{remark}

\section{Numerical Experiments}\label{SecExperiments}

Here we compare the performances of the presented wireless edge learning schemes for the task of image classification. We run experiments on the MNIST dataset \cite{LeCunMNIST} with $60000$ training and $10000$ test samples, and train a single layer neural network with $d=7850$ parameters utilizing ADAM optimizer \cite{ADAMDC}. Throughout the experiments, we consider $\sigma^2=1$, and $s= \left\lceil d/20 \right\rceil$ parallel subchannels, which results in $N = 10$ for ESA-DSGD and ECESA-DSGD; and for any $\tilde{s}$ of the CA-DSGD scheme, we set the sparsity level to $k = \left\lfloor \tilde{s}/2.5 \right\rfloor$. For a fair comparison between the analog DSGD schemes, we set $\lambda^{\rm{x}} (t) = \lambda, \forall {\rm{x}} \in \{ {\rm{e}}, {\rm{v}}, {\rm{c}} \}$, and for average transmit power $\bar{P}^n(t)$ at the $n$-th time slot, $n \in [N]$, of the $t$-th iteration, $t \in [T]$, we calculate values of $\gamma^{{\rm{e}}, n}_m (t)$, $\gamma^{{\rm{v}}, n}_m (t)$ and $\gamma^{{\rm{c}}, n}_m (t)$ for ESA-DSGD, ECESA-DSGD and CA-DSGD through \eqref{GammaESA}, \eqref{GammaECESA} and \eqref{GammaCA}, respectively. Also, we consider $\bar{P}^n(t) = \bar{P}$, $\forall n, t$, for the analog schemes, and $\bar{P}_{m^*(t)}(t) = \bar{P}$, $\forall t$, for the digital schemes. The performance is measured as the accuracy with respect to the test data samples, called \textit{test accuracy}, versus the normalized time $Nt$.

We consider two scenarios to model the data distribution across the devices: in \textit{IID data distribution}, $B$ randomly selected training data samples are assigned to each device at the beginning of training; while in \textit{non-IID data distribution}, each device has $B$ training data samples, where half of them are selected at random from only one class/label; that is, for each device, we first select two classes/labels at random, and then randomly select $B/2$ data samples from each of the two classes/labels. At each iteration, devices use all the $B$ local data samples to compute their gradient estimates, i.e., the batch size is equal to the size of the local datasets.


For numerical comparison, we also consider the \textit{error-free shared link} approach, where at each time slot the PS receives the average of the actual gradient estimates computed by the devices, $\frac{1}{M} \sum\nolimits_{m=1}^{M} \boldsymbol{g}_m \left(\boldsymbol{\theta}_t \right)$, in a noiseless fashion, and updates the parameters based on this error-free observation. We note that, for the error-free shared link approach, we have $N=1$. We consider three alternative digital schemes employing sparse binary compression \cite{DCSattlerSparseBinary}, QSGD \cite{DCAlistarhQSGD} and SignSGD \cite{SignSGDBernstein} algorithms for gradient compression, respectively. When we refer to D-DSGD it refers to using the sparse binary compression technique. For a fair comparison, we apply QSGD and SignSGD to a limited number of gradient entries such that the final number of bits does not exceed the capacity of the underlying fading MAC. To be more precise, considering the device scheduling policy given in \eqref{WorkSelectionDig}, $q^{\rm{S}} (t)$ and $q^{\rm{Q}} (t)$ gradient entries with highest magnitudes are selected for transmission for SignSGD and QSGD, respectively, while all the other entries are set to zero. With the SignSGD algorithm \cite{SignSGDBernstein}, the scheduled device transmits only the signs of the $q^{\rm{S}} (t)$ selected entries, and a total of
\begin{align}\label{SignSGDDeliveredBits}
r^{\rm{S}} (t) = \log_2 \binom{d}{q^{\rm{S}}(t)} + {q^{\rm{S}}(t)} \mbox{ bits}, \quad \forall t ,   
\end{align}
are required to send the sign of each selected entry, as well as their locations, and $q^{\rm{S}}(t)$ is set as the largest integer satisfying $r^{\rm{S}} (t) \le R(t)$. 
With the QSGD algorithm \cite{DCAlistarhQSGD}, the scheduled device transmits a quantized version of each of the $q^{\rm{Q}} (t)$ selected entries with a quantization level of $2^{l^{\rm{Q}}}$, the $l_2$-norm of the resultant vector with $q^{\rm{Q}} (t)$ non-zero entries, and the locations of the non-zero entries. Thus, a total of
\begin{align}\label{QSGDDeliveredBits}
r^{\rm{Q}} (t) = 32 + \log_2 \binom{d}{q^{\rm{Q}} (t)} + (1+l^{\rm{Q}}) {q^{\rm{Q}} (t)} \mbox{ bits}, \quad \forall t,    
\end{align}
are sent over the wireless fading MAC, where ${q^{\rm{Q}} (t)}$ is set as the largest integer satisfying $r^{\rm{Q}} (t) \le R_t$. Here we consider a quantization level $l^{\rm{Q}}=2$ for QSGD.

We further consider the OD-DSGD scheme, where each device has access to $\left\lfloor {s/M} \right\rfloor$ distinct subchannels to perform digital transmission without interfering with other devices. Due to symmetry across devices, we allocate subchannels $(m-1) \left\lfloor {s/M} \right\rfloor +1$ to $m \left\lfloor {s/M} \right\rfloor$ to device $m$, $m \in [M]$. Similarly to the D-DSGD scheme, we use the capacity upperbound to determine the number of bits each user can convey to the PS at each iteration. This bound is computed by waterfilling across the $\left\lfloor {s/M} \right\rfloor$ channels available to each device as in \eqref{CapacityDDSGD}-\eqref{CapacityDDSGDRt}. 
We use sparse binary compression with the OD-DSGD scheme as well with the sparsity level $q_m^{\rm{O}} (t)$ set as the largest integer satisfying $\log_2 \binom{d}{q^{\rm{O}}_m (t)} + 33 \le R_m^{{\rm{O}}} (t)$, $\forall t$.

\begin{figure*}[t!]
\centering
\begin{subfigure}{.5\textwidth}
  \centering
  \includegraphics[scale=0.575,trim={20pt 7pt 45pt 40pt},clip]{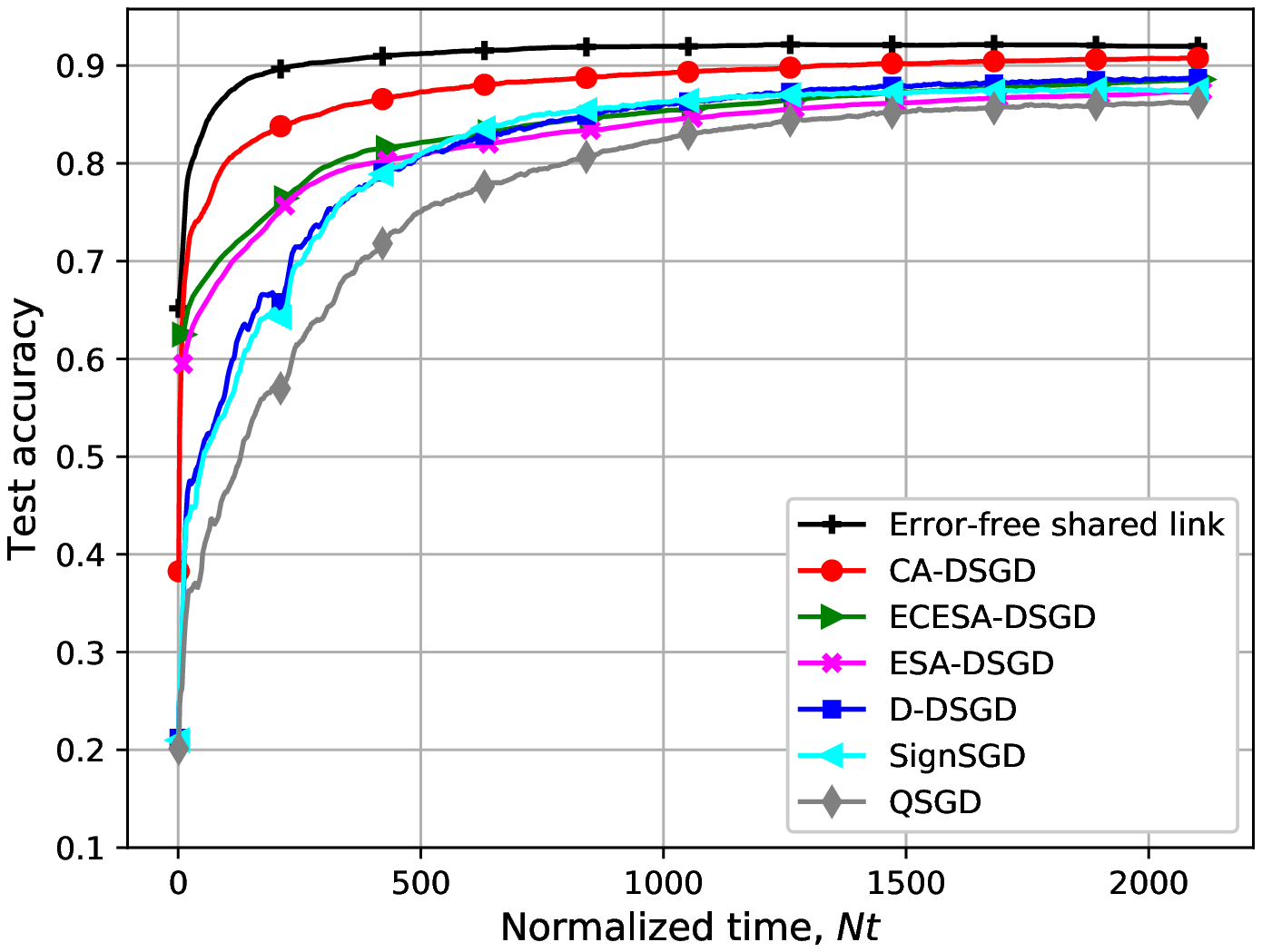}
  \caption{IID data distribution}
  \label{Figure_analog_digital_IID}
\end{subfigure}%
\begin{subfigure}{.5\textwidth}
  \centering
  \includegraphics[scale=0.575,trim={20pt 7pt 45pt 40pt},clip]{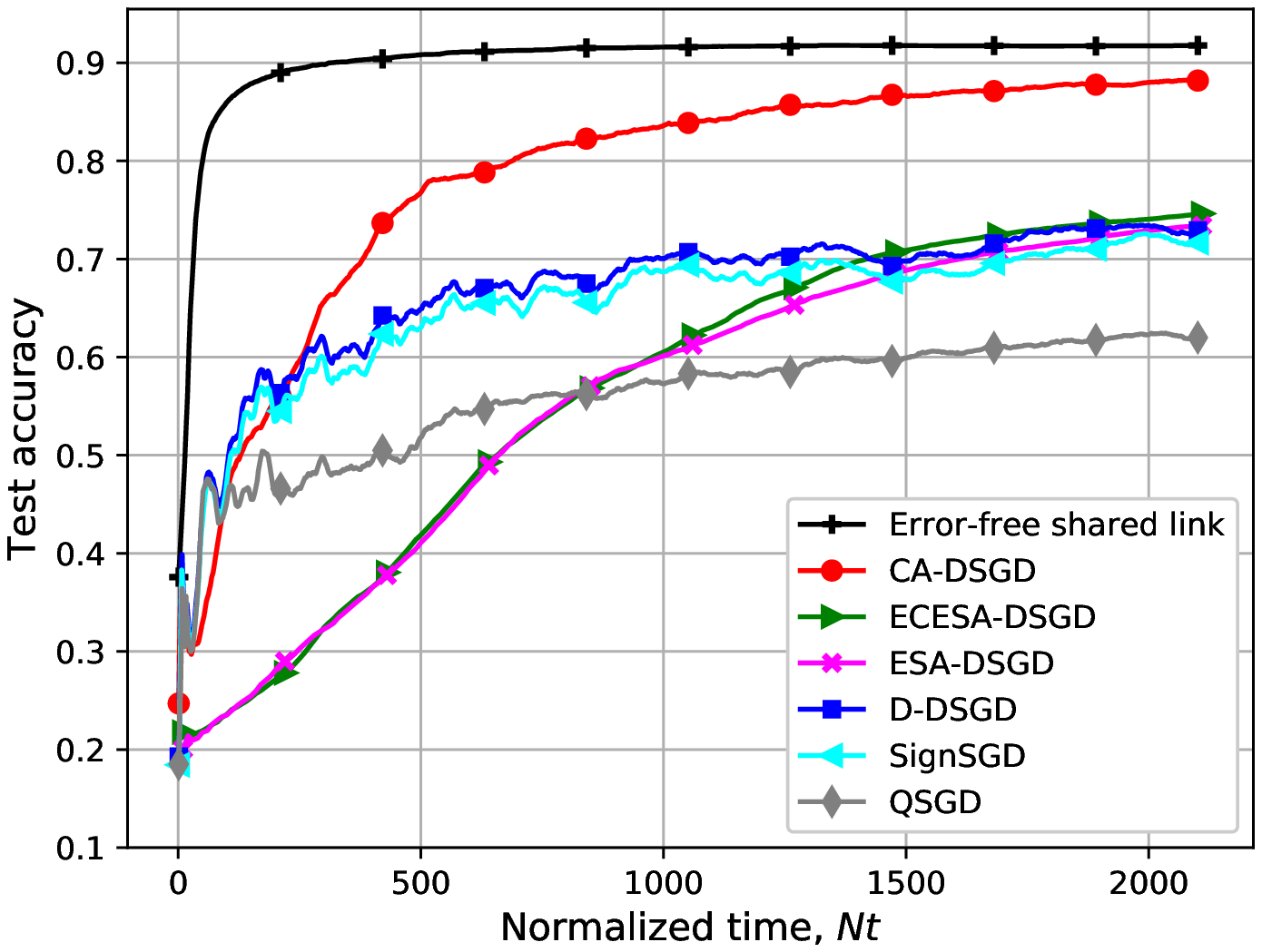}
  \caption{Non-IID data distribution}
  \label{Figure_analog_digital_nonIID}
\end{subfigure}
\caption{Test accuracy of different algorithms for IID and non-IID data distribution scenarios for $M=25$, $B=1000$ and $\bar{P} = 20$.}
\label{Figure_analog_digital_IIDnonIID}
\end{figure*}

In Fig. \ref{Figure_analog_digital_IIDnonIID}, we compare the performances of different analog and digital schemes for IID and non-IID data distribution scenarios, for $M=25$ devices, $B=1000$ training data samples and average transmit power constraint $\bar{P} = 20$. We consider $\tilde{s}=2s=d/10$, i.e., $N=1$ for CA-DSGD, and we set the threshold value to $\lambda = 10^{-3}$. Observe that for both IID and non-IID data distribution cases CA-DSGD outperforms all other analog and digital schemes with the improvement substantially larger for non-IID data distribution, which shows its robustness to bias in the data distribution. CA-DSGD has a smaller convergence speed in non-IID case which is due to the reduction in the similarity of the sparsity patterns of the gradients across devices, although it does converge much faster and to a much higher accuracy level compared to the other schemes under consideration. The gap between the error-free shared link approach and CA-DSGD is relatively small for the IID case, and the final test accuracies of the two approaches are also similar for the non-IID case, although CA-DSGD converges more slowly in this case. Unlike the digital schemes, CA-DSGD benefits from the superposition property of the underlying wireless MAC by aligning the transmit powers to dominate the noise. We further highlight that the main reasons for the degradation of the ESA-DSGD over CA-DSGD are i) scheduling gradient entries for transmission only based on the channel gains; ii) transmitting the entire gradient vectors of relatively huge dimensions (compared to the channel bandwidth); iii) ignoring the gradient entries which have not been transmitted due to the poor conditions of their corresponding channels. We note that ECESA-DSGD resolves the last issue by utilizing error accumulation technique, which provides some gains with respect to ESA-DSGD, but we observe that better scheduling of the gradient transmissions and more efficient utilization of the bandwidth through linear projection provide significant gains, especially for the non-IID case, where the performances of ESA-DSGD and ECESA-DSGD significantly degrade in terms of the test accuracy, as well as the convergence speed. Also, D-DSGD provides a better accuracy than SignSGD and QSGD for both data distribution scenarios, and the performance of all the digital schemes under consideration deteriorate substantially for the non-IID case; this performance loss is more severe for the QSGD scheme.

\begin{figure*}[t!]
\centering
\begin{subfigure}{.5\textwidth}
  \centering
  \includegraphics[scale=0.575,trim={20pt 7pt 45pt 40pt},clip]{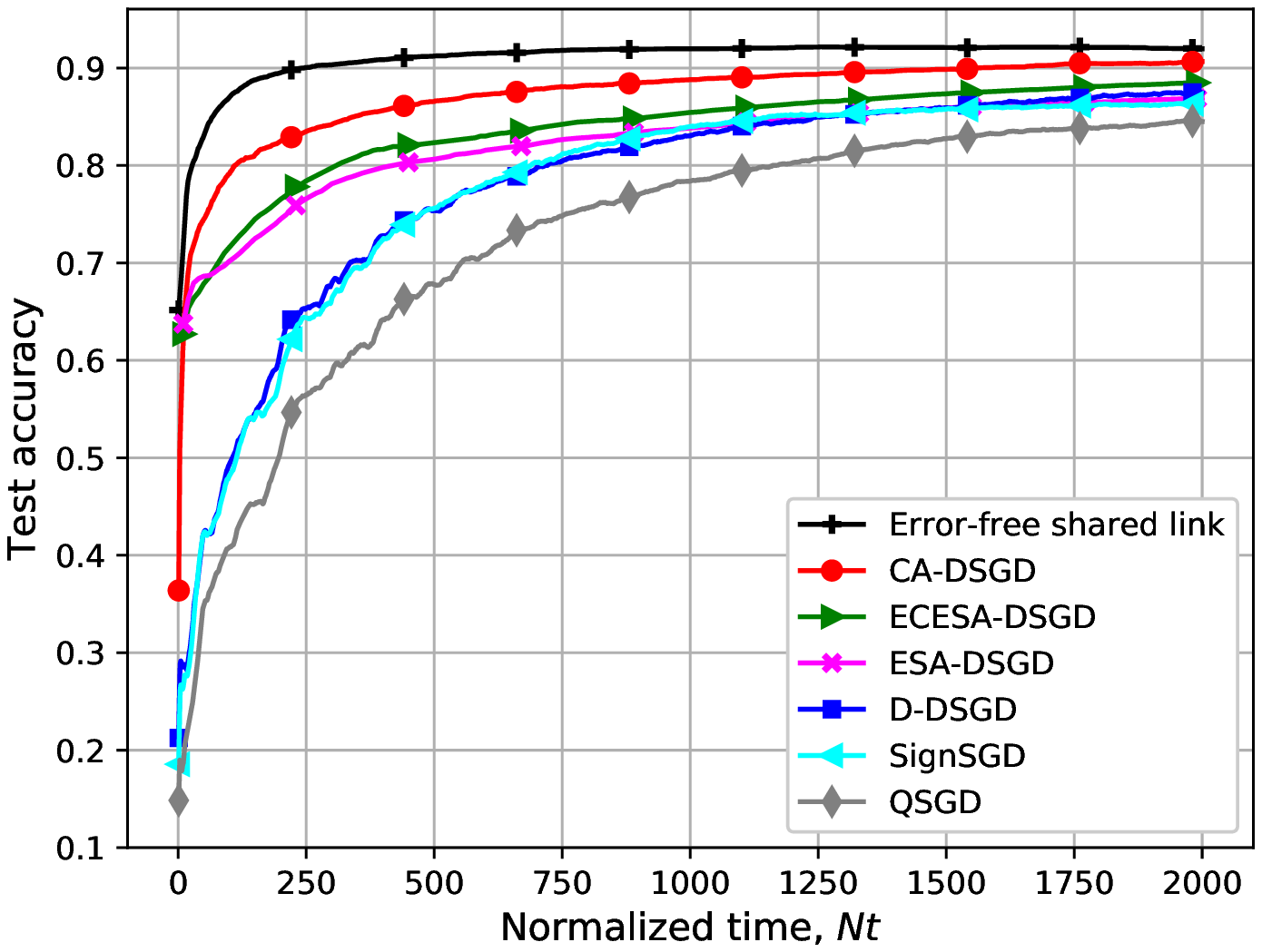}
  \caption{$\bar{P} = 10$}
  \label{CA_ESA_AccESA_Dig_P1P2_P1}
\end{subfigure}%
\begin{subfigure}{.5\textwidth}
  \centering
  \includegraphics[scale=0.575,trim={20pt 7pt 45pt 40pt},clip]{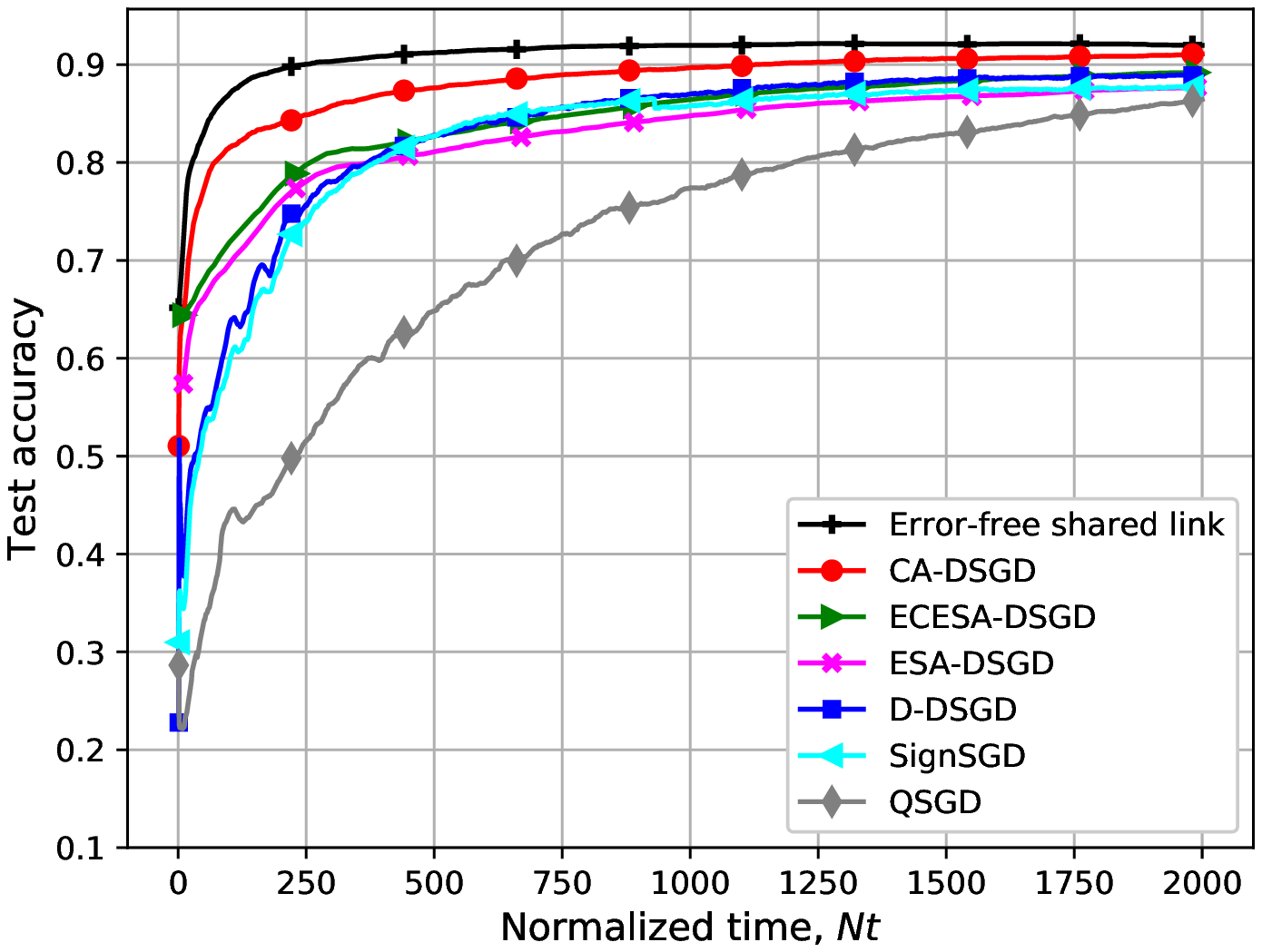}
  \caption{$\bar{P} = 30$}
  \label{CA_ESA_AccESA_Dig_P1P2_P2}
\end{subfigure}
\caption{Test accuracy of different algorithms for two different $\bar{P}$ values, $\bar{P} \in \{ 10, 30 \}$, when $M=25$ and $B=1000$.}
\label{CA_ESA_AccESA_Dig_P1P2}
\end{figure*}

In the following experiments, we only consider IID data distribution. In Fig. \ref{CA_ESA_AccESA_Dig_P1P2}, we compare the performances of different analog and digital algorithms for two different average transmit power values $\bar{P} = 10$ and $\bar{P} = 30$. We consider $M=25$ and $B=10$, and we set $\tilde{s}=2s =d/10$, i.e., $N = 1$ for CA-DSGD, and $\lambda = 5 \times 10^{-3}$. As it can be seen, CA-DSGD continues to outperform all the other schemes, with a relatively small gap to the error-free shared link approach. 
By comparing Figures \ref{CA_ESA_AccESA_Dig_P1P2_P1} and \ref{CA_ESA_AccESA_Dig_P1P2_P2}, it can be seen that the performances of all the schemes improve with $\bar{P}$, but the improvement is more significant for the digital schemes in terms of both the accuracy and the convergence speed, except QSGD which only improves in terms of accuracy. This shows that the analog schemes are less sensitive to a reduction in the average transmit power than the digital ones. This is because the signal-superposition property of the wireless MAC aligns the transmission power of all the transmitting devices in the case of analog transmission, which provides stronger protection against noise collectively, as the system continues to operate in a relatively high effective signal-to-noise ratio (SNR) regime despite some reduction in the transmission power of individual devices. 

\begin{figure*}[t!]
\centering
\begin{subfigure}{.5\textwidth}
  \centering
  \includegraphics[scale=0.575,trim={20pt 7pt 45pt 40pt},clip]{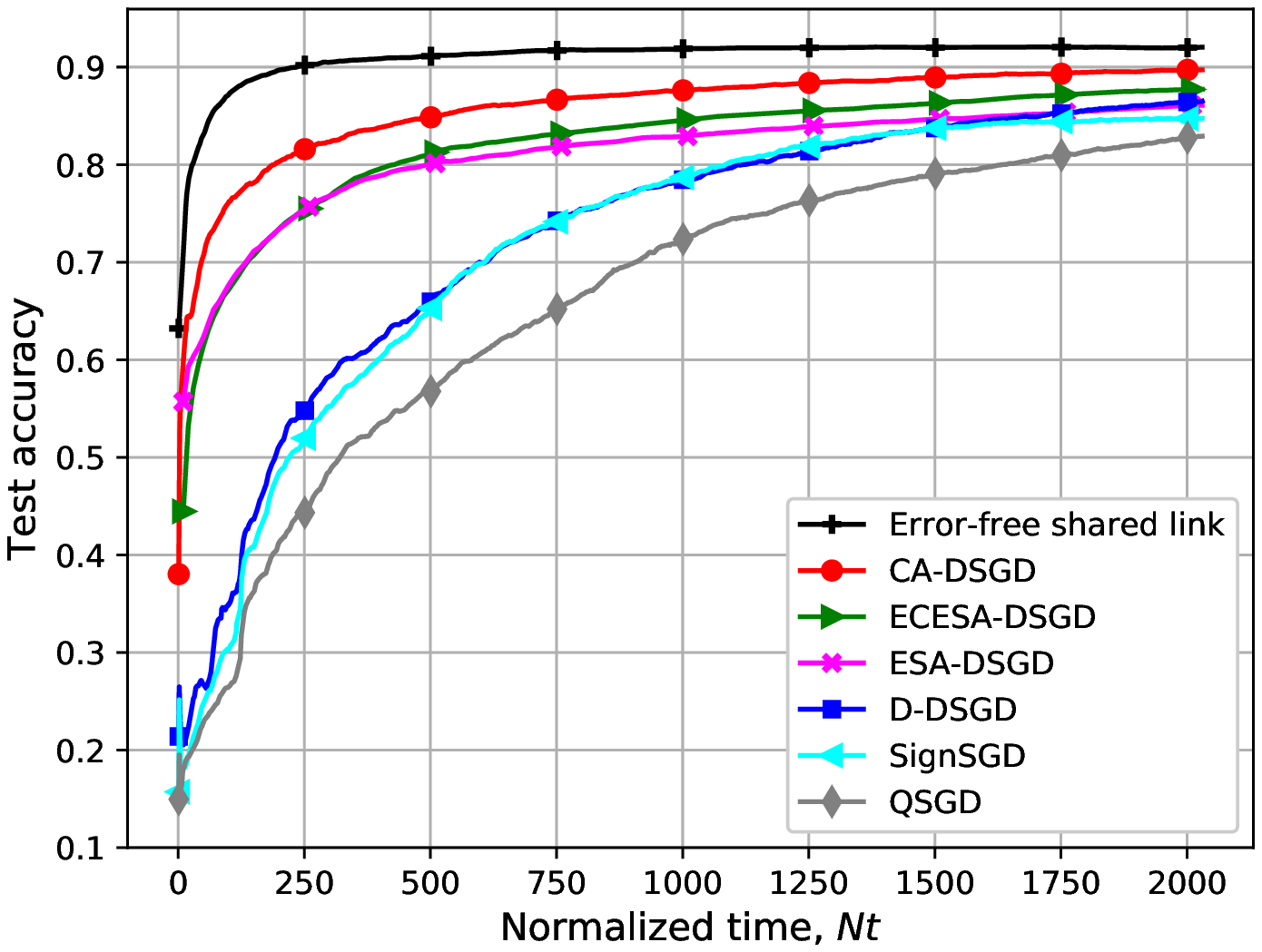}
  \caption{$(M,B) = (15, 2000)$}
  \label{CA_ESA_AccESA_Dig_M1M2_M1}
\end{subfigure}%
\begin{subfigure}{.5\textwidth}
  \centering
  \includegraphics[scale=0.575,trim={20pt 7pt 45pt 40pt},clip]{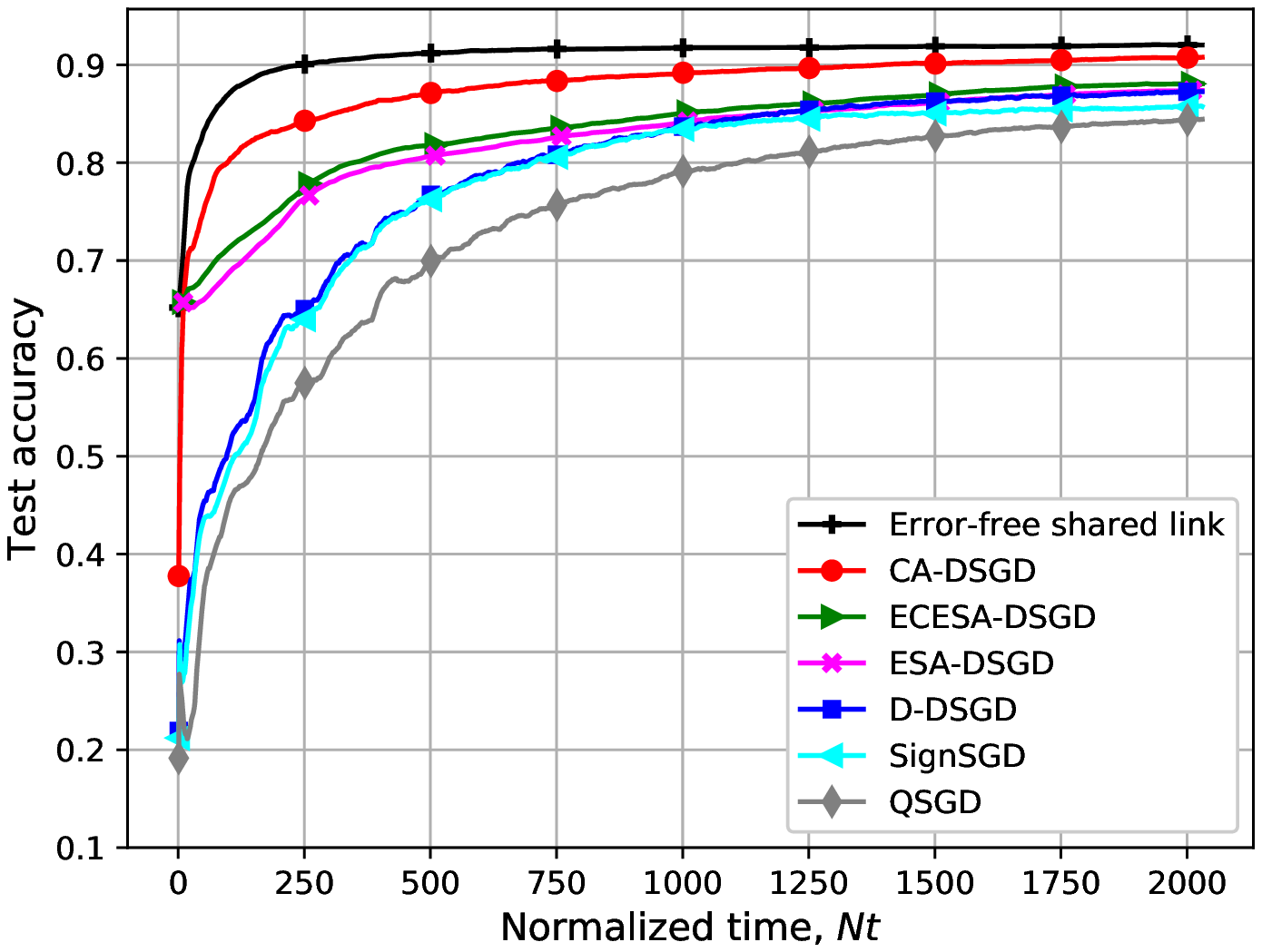}
  \caption{$(M,B) = (30, 1000)$}
  \label{CA_ESA_AccESA_Dig_M1M2_M2}
\end{subfigure}
\caption{Test accuracy of different algorithms different $(M,B)$ pairs when $MB$ is fixed, and $\bar{P} =10$.}
\label{CA_ESA_AccESA_Dig_M1M2}
\end{figure*}

In Fig. \ref{CA_ESA_AccESA_Dig_M1M2}, we compare the performances of analog and digital schemes for different $(M,B)$ pairs, $(M, B) \in \{ (15, 2000), (30, 1000) \}$, both having the same size of training data in total. We consider $\bar{P} =10$, and we set $\lambda = 5 \times 10^{-3}$, and $\tilde{s} = 2s = d / 10$, which results in $N = 1$ for CA-DSGD. It is again evident that CA-DSGD outperforms all the other schemes with the improvement over ESA-DSGD and ECESA-DSGD more noticeable for the higher $M$ value. As it can be seen, the performances of the analog schemes improve with $M$, since increasing $M$ provides additional power introduced by each device and increases the robustness of the estimation against noise. We note that this improvement is larger for CA-DSGD, which is due to the more efficient utilization of the gradient estimates computed by the devices. Digital schemes also gain from increasing $M$, which is due to the additional power allocated to the selected device, as the devices are less frequently scheduled for transmission. We note that the superiority of the ECESA-DSGD over ESA-DSGD reduces with $M$, which shows that error accumulation is less effective for higher $M$ values when $MB$ is fixed. This is because for larger $M$, the chance of receiving more estimates for each entry of the actual gradient vector is higher (each gradient entry is estimated more accurately at the PS), and it is less likely that no estimate of any gradient entry is received by the PS. Accordingly, the benefit of error accumulation becomes less significant for higher number of devices.

\begin{figure}[t!]
\centering
\includegraphics[scale=0.71,trim={20pt 7pt 45pt 40pt},clip]{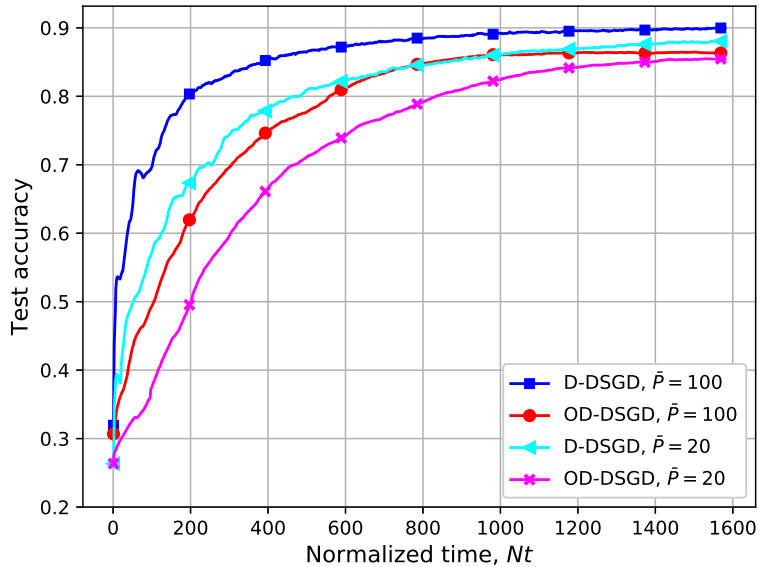}
\caption{Test accuracies of D-DSGD and OD-DSGD schemes for different $\bar{P}$ values, $\bar{P} \in \{20, 100\}$, when $M=25$ and $B=1000$.}
\label{Figure_digital_orthogonal}
\end{figure}

In Fig. \ref{Figure_digital_orthogonal}, we compare the performance of D-DSGD with that of OD-DSGD for different $\bar{P}$ values, $\bar{P} \in \{ 20, 100\}$, when $M=25$ and $B=1000$. For both power values we observe that D-DSGD significantly outperforms OD-DSGD in terms of accuracy and convergence speed, while the superiority is more highlighted for the higher $\bar{P}$ value. This shows that opportunistically allocating all the available bandwidth to only a single device is better than sharing it equally among all the devices which indicates that it is better to receive an accurate gradient estimate from a single device at each iteration, instead of receiving coarse estimates from all the devices. This improvement is more significant when $\bar{P}$ increases. 

\begin{figure}[t!]
\centering
\includegraphics[scale=0.75,trim={20pt 7pt 45pt 40pt},clip]{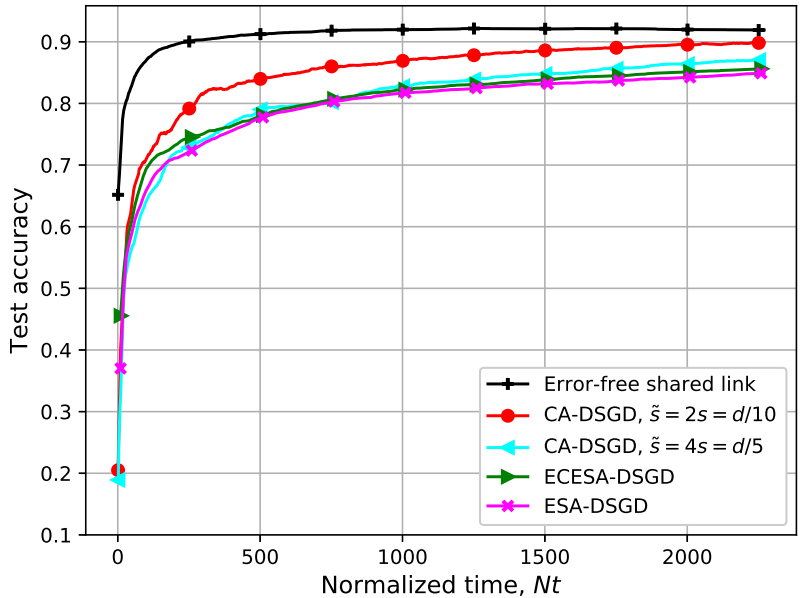}
\caption{Test accuracies of different analog schemes with different $\tilde{s}$ values, $\tilde{s} \in \{ 2s, 4s \}$, for the CA-DSGD scheme, when $M=25$, $B=1000$, $\bar{P} =1$.}
\label{CA_ESA_AccESA_s1s2}
\end{figure}

In Fig. \ref{CA_ESA_AccESA_s1s2} we investigate the impact of $\tilde{s}$ on the performance of CA-DSGD. We consider $\tilde{s} \in \{ 2s, 4s \} = \{ d/10, d/5 \}$ for CA-DSGD, in which $\tilde{s} = 2s$ and $\tilde{s} = 4s$ are equivalent to $N=1$ and $N=2$, respectively. We have $M=15$, $B=1000$ and $\bar{P}=1$, ans we set $\lambda = 5 \times 10^{-3}$. We highlight the superiority of the CA-DSGD scheme for both $\tilde{s}$ values under consideration over ESA-DSGD. The ECESA-DSGD scheme, which is equivalent to CA-DSGD for $\tilde{s} = d$, also outperforms ESA-DSGD slightly. However, as it can be seen, the performance of CA-DSGD degrades as $\tilde{s}$ increases, which indicates that transmitting more sparse versions of the gradient estimates while using the available channel bandwidth for further iterations results in a higher accuracy. The flexibility in choosing the dimension of the transmitted gradient estimates makes the proposed CA-DSGD scheme particularly compelling for learning at the wireless edge under strict bandwidth limit.

\begin{figure}[t!]
\centering
\includegraphics[scale=0.75,trim={20pt 7pt 45pt 40pt},clip]{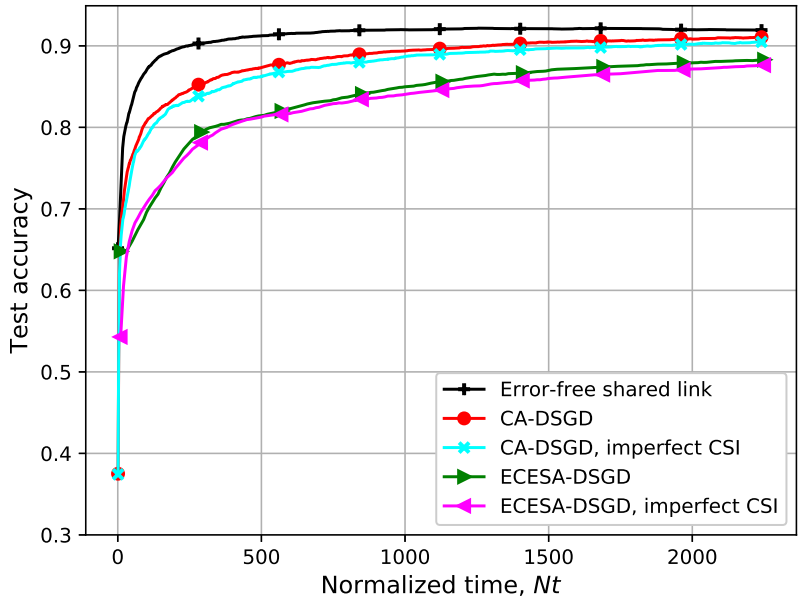}
\caption{Test accuracies of CA-DSGD and ECESA-DSGD schemes under imperfect CSI $\hat{h}_{m,i}^n (t) = {h}_{m,i}^n (t) + \tilde{h}_{m,i}^n (t)$, $\forall m,n,i,t$, with $\tilde{h}_{m,i}^n (t)$ i.i.d. according to $\mathcal{CN}(0,1)$, when $M=25$, $B=1000$ and $\bar{P} = 10$.}
\label{Figure_analog_imperfect_CSI}
\end{figure}

In Fig. \ref{Figure_analog_imperfect_CSI}, we consider the impact of imperfect CSI on the performance of analog schemes CA-DSGD and ECESA-DSGD for $M=25$, $B=1000$ and $\bar{P} = 10$. We set $\lambda = 5 \times 10^{-3}$, and $\tilde{s} = 2s = d / 10$, which results in $N = 1$ for CA-DSGD. We assume a noisy CSI at device $m$ given by $\hat{h}_{m,i}^n (t) = {h}_{m,i}^n (t) + \tilde{h}_{m,i}^n (t)$, $\forall m,n,i,t$, where $\tilde{h}_{m,i}^n (t)$ is i.i.d. according to $\mathcal{CN}(0,1)$, i.e., a complex normal random variable with the same variance as the actual channel gain ${h}_{m,i}^n (t)$. We note that all the processing at the devices, such as power allocation, finding the set $\mathcal{M}^n_i (t)$, and obtaining $\gamma^{{\rm{v}}, n}_m (t)$ and $\gamma^{{\rm{c}}, n}_m (t)$, and consequently $\bar{\gamma}^{{\rm{v}}, n} (t)$ and $\bar{\gamma}^{{\rm{c}}, n} (t)$ for CA-DSGD and ECESA-DSGD, respectively, are performed based on the imperfect CSI $\hat{h}_{m,i}^n (t)$, $\forall m,n,t$. As it can be seen, both the CA-DSGD and ECESA-DSGD are robust against the imperfect CSI, and their performance loss is negligible. To be more precise, after 2250 time slots (2250 SGD iterations for CA-DSGD and 225 SGD iterations for ECESA-DSGD), the final test accuracy reduction for CA-DSGD due to the imperfect CSI is $0.67\%$, and that of ECESA-DSGD is $0.76\%$. We highlight that, with imperfect CSI, even though the users will allocate higher or lower power to each subchannel than the optimal one, the cumulative effect becomes negligible since these variations across users are averaged out thanks to the superposition property.

\section{Conclusions}\label{SecConc}

We have studied FL at the wireless edge, where $M$ devices with limited transmit power and datasets communicate with the PS over a bandwidth-limited fading MAC to minimize a loss function by performing DSGD. The PS updates the parameter vector, and shares it with the devices over an error-free shared link. We first presented a digital approach that treats computation and communication separately. At each iteration of the proposed D-DSGD scheme, one device is selected depending on the channel states, and the selected device first quantizes its gradient estimate, and transmits the quantized bits to the PS using a capacity-achieving channel code. Then we studied an alternative analog transmission approach, which does not employ quantization or channel coding, and exploits the superposition property of the wireless MAC, rather than orthogonalizing the transmissions from different devices. We have proposed the CA-DSGD scheme, where each device employs gradient sparsification with error accumulation followed by linear projection to reduce the typically very large parameter vector dimension to the limited channel bandwidth. We have also designed a power allocation scheme to align the received vectors at the PS while satisfying the average power constraints at the devices. The CA-DSGD scheme allows a much more efficient use of the limited channel bandwidth, and benefits from the ``beamforming effect'' thanks to the similarity in the patterns of the gradient estimates across devices. The impact of various system parameters on the performance is studied numerically considering MNIST classification across edge devices as an example. Numerical results show that CA-DSGD outperforms D-DSGD and other state-of-the-art analog schemes consistently, while this improvement is even more significant for the non-IID data distribution scenario. 

\bibliographystyle{IEEEtran}
\bibliography{Report}

\end{document}